# Ultrasensitive Real-Time Detection of SARS-CoV-2 Proteins with Arrays of Biofunctionalized Graphene Field-Effect Transistors

*Hamid Reza Rasouli[1], David Kaiser[1], Ghazaleh Eshaghi[1], Marco Reinhard[2], Alexander Rolapp[2], Dominik Gary[3], Tobias Fischer[3], Christof Neumann[1], Thomas Weimann[4], Katrin Frankenfeld[3], Michael Meister[2], Andrey Turchanin[1,5]\**

[1]Institute of Physical Chemistry, Friedrich Schiller University Jena, 07743 Jena, Germany

[2]IMMS Institut für Mikroelektronik- und Mechatronik-Systeme gemeinnützige GmbH (IMMS GmbH), 99099 Erfurt, Germany

[3]fzmb GmbH, Forschungszentrum für Medizintechnik und Biotechnologie, 99947 Bad Langensalza, Germany

[4]Physikalisch-Technische Bundesanstalt (PTB), 38116 Braunschweig, Germany

[5]Jena Center for Soft Matter (JCSM), 07743 Jena, Germany

**Corresponding Author:** Andrey Turchanin (andrey.turchanin@uni-jena.de)

**Graphical Abstract**

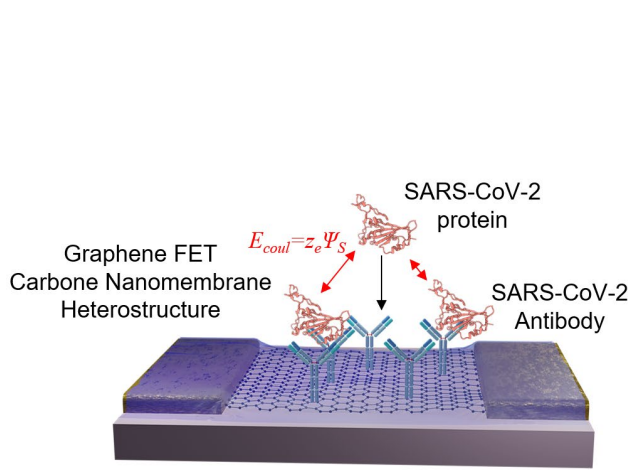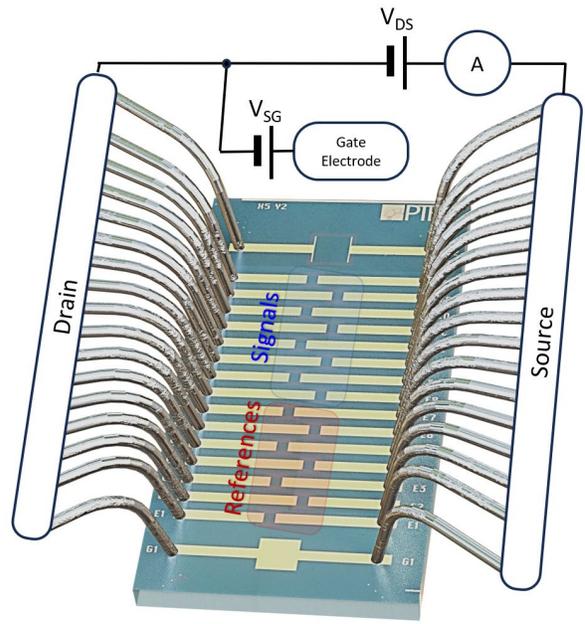

Parallel real-time monitoring

**Abstract**

With the growing interest in graphene field-effect transistors (GFETs) for biosensing applications, there is a strong demand for strategies enabling flexible and multiplexed biofunctionalization, as well as highly parallel, real-time electronic readout integrated with microfluidic control. Here we present a methodology that addresses these challenges by enabling real-time, parallel monitoring of multiple GFETs integrated on a single microfabricated chip within an automated electronic and microfluidic platform. We demonstrate the capabilities of this approach through ultrasensitive detection of the SARS-CoV-2 spike (S) and nucleocapsid (N) proteins. GFET chips are functionalized *via* van der Waals assembly using 1 nm-thick molecular two-dimensional (2D) materials – carbon nanomembranes – which enable multiplexed biofunctionalization. The chips are integrated into a custom-developed microelectronic and microfluidic system that allows parallel, real-time, and automated measurements of 15 GFETs. We present *in situ* biofunctionalization of the GFETs with antibodies, followed by highly specific detection of the S- and N-proteins with limits of detection down to 10 aM and a dynamic range spanning four orders of magnitude. Owing to its versatility, the presented methodology is readily adaptable for sensing a wide range of biological and chemical targets.

**Keywords:**

Biosensors, 2D Materials, Graphene Field-Effect Transistors, Molecular Nanotechnology, SARS-CoV-2

**Introduction**

The outbreak of the coronavirus SARS-CoV-2 has underscored the critical importance of early and accurate viral detection to prevent the escalation of global health crises.[1] Despite the accuracy and high sensitivity of established detection methods such as reverse transcription polymerase chain reaction (RT-PCR), their complexity and reliance on specialized laboratory conditions make them unsuitable for point-of-care (PoC) applications and continuous, real-time biosensing.[2-4] These drawbacks have prompted a shift towards the development of portable, single-chip biosensors.[5, 6] The development of nanotechnology and its integration into biosensing platforms has opened new avenues for virus detection, enabling faster, sensitive, and PoC diagnostic solutions.[7-9] In particular, graphene field-effect transistors (GFETs) are renowned for their extraordinary electronic properties, high sensitivity, and biocompatibility, offering the capability for direct detection of biomolecules without requiring complex sample preparation or amplification procedures.[10-13] While scalable, multiplexed GFET arrays are advancing high-throughput screening,[14, 15] their underlying electronic architecture limits data fidelity and temporal resolution. So far, most systems have employed time-division multiplexing,[16-23] a sequential process that causes the total array measurement time to scale directly with the number of sensors and introduces undesired temporal offsets between signal and reference channels.

The integration of GFETs with other nanomaterials continues to enhance their performance and applicability.[24-28] However, functionalization of pristine graphene surface with bioreceptors presents significant challenges due to graphene's chemically inert nature. While covalent methods can disrupt graphene's electronic properties, non-covalent functionalization preserves them, making it attractive for sensing applications.[29] Nonetheless, non-covalent methods may suffer from inherently weaker binding and therefore reduced stability, posing the challenge of maintaining a balance between functionalization robustness and electronic property preservation in graphene-based biosensors. In this respect, the use of carbon nanomembranes (CNMs) and graphene van der Waals (vdW) heterostructures [24, 26, 30] in GFETs provides robust binding of biorecognition elements such as aptamers[27] and antibodies[31] while preserving the intrinsic properties of the underlying graphene. Notably, 1-nm-thick CNM enables high proximity of the biomolecules to the graphene surface without compromising the sensor's sensitivity.[27]

Besides biofunctionalization of GFETs, another challenge in their implementation as biosensors is the electrical charge screening of the target biomolecules under physiological ionic-strength conditions.[32-34] Although decreasing the ionic strength of the buffer, which results in an increase of the Debye length ($\lambda_D$),[35-37] is a straightforward strategy to mitigate charge screening, it requires careful consideration to preserve biomolecules' structure and their binding activity.[38-40] Moreover, lowering the ionic strength by diluting the buffer comes at the cost of simultaneously reducing the analyte concentration, which may hinder the detection of low-abundance targets.[41] In practice, physiological buffer conditions are often optimal for maintaining biomolecular binding affinity, yet the high ionic strength in such buffers strongly screens the charge of bound analytes. One effective strategy is therefore to perform the binding step in physiological ionic strength buffers and subsequently exchange the solution with a diluted buffer for the electrical measurement, thereby extending the Debye length and improving the sensor's sensitivity.[18, 35, 42] Reliable implementation of this approach requires controlled buffer exchange and precise recording of electrical characteristics, a process that can be facilitated by automated microfluidic systems to minimize artifacts and reduce the risk of misinterpreting biosensor performance.

By addressing the challenges of the biofunctionalization stability, the charge screening in physiological conditions, and the need for parallel real-time data acquisition in GFET array biosensors, we introduce a methodology for ultra-sensitive detection of SARS-CoV-2 proteins, namely receptor-binding domain spike (S-protein) and nucleocapsid (N-protein) proteins, using GFETs functionalized with CNMs *via* van der Waals assembly. This functionalization enables covalent immobilization of antibodies for specific detection of the target proteins without compromising the graphene's electronic properties. A distinctive aspect of our approach is the use of a dedicated source-measure unit (SMU) for each of the 15 GFETs in the sensor array, enabling simultaneous acquisition of transfer curves for all the devices in liquid. This readout architecture allows parallel real-time monitoring of the charge neutrality point ($V_{CNP}$) across all the devices without temporal offset. The GFET array includes built-in reference devices (functionalized with non-complementary antibodies), enabling differential sensorgram analysis of $V_{CNP}$ in real time. The combination of this measurement setup with an automated microfluidic system enables continuous monitoring of $V_{CNPs}$ during controlled buffer exchange and target injection, thereby facilitating binding recognition under physiological ionic strength buffers and subsequent measurements in diluted buffers. Moreover, we calculate the sensor response using an analytical model that considers the charge of the antibodies and their electrostatic interaction with the targets. Based on this methodology, we have achieved limits of detection (LOD) of 10 attomolar for the S-protein and 100 attomolar for the N-protein, and demonstrated the capability for dual detection of these proteins using a single GFET chip.

**Results and Discussion**

*Biofunctionalization of GFETs and their Integration with Readout Electronics and Microfluidics*

As presented in Figure 1a, the biofunctionalization of the GFET surface is achieved *via* the vdW heterostructure assembly with azide-functionalized CNM (N$_3$-CNM)[27] (see Methods for details).

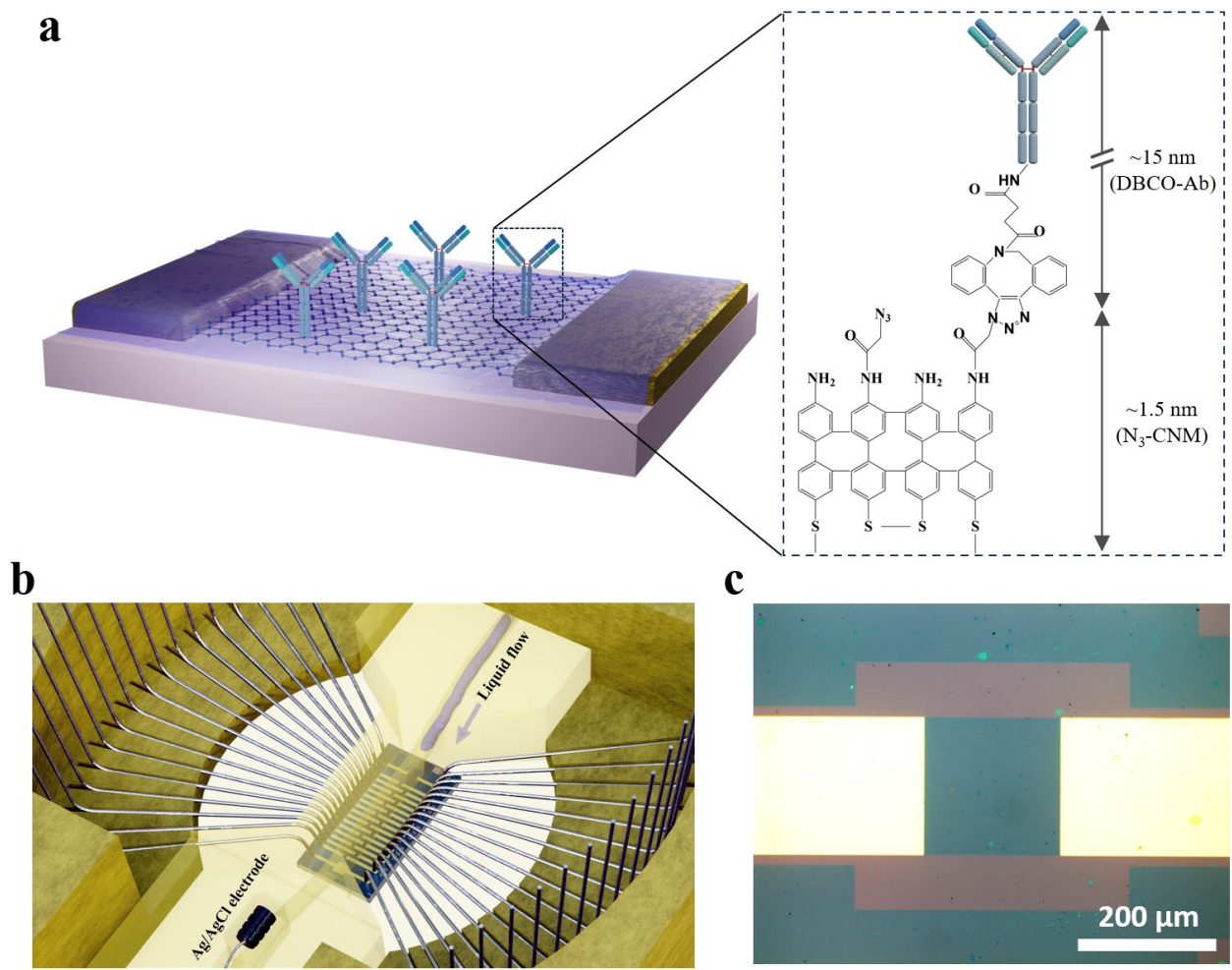

**Figure 1**. Biofunctionalization of GFETs and their integration into the microfluidic system. a) Schematic illustration of a $N_3$-CNM/graphene FET device highlighting the immobilization of SARS-CoV-2 antibodies on the $N_3$-CNM surface. The inset details the CNM's chemical structure with azide linkers ($N_3$-CNM),[31] enabling immobilization of the DBCO-labeled antibody (DBCO-Ab) *via* click chemistry. b) An enlarged schematic view of the microfluidic cell integrated with a GFET chip, where each device is interfaced with two probe needles for source and drain connections. c) Optical image of a typical GFET device microfabricated on a $SiO_2$/Si substrate.

The $N_3$-CNM serves as a 2D ultrathin molecular interposer for the immobilization of SARS-CoV-2 antibodies.[31] This immobilization is achieved through click chemistry reactions between the azide groups on the CNM and dibenzocyclooctyne (DBCO) labeled antibodies (DBCO-Ab). Note that the estimated nominal thickness of the densely packed DBCO-Ab layer, as shown in Figure 1a, is ~15 nm, while its real effective thickness in physiological PBS-P buffer was experimentally determined to be ~4 nm.[31]

The antibody-functionalized GFET chip is subsequently integrated into the microfluidic cell (Figure 1b) where the central area of the chip is placed under the fluidic channel with a cross-sectional area of 1×1 mm². The flow and timing of the liquid injections in the channel are controlled by an automated microfluidic system (see Figure S1 in the Supporting Information for the entire experimental setup), enabling programmable and reproducible delivery of buffer, antibody, and antigen solutions throughout the experiments. The GFET chip consists of 15 individual devices, each with an active area of 200 × 200 μm² microfabricated on an oxidized Si wafer (Figure 1c). Each device is contacted by two probe needles (source and drain) positioned on the gold electrodes outside of the microfluid channel. The solution gate modulation of all the devices is realized through an Ag/AgCl electrode integrated into the microfluid channel, with a 10 mV bias applied between source and drain ($V_{ds}$). In comparison to previously reported GFET array systems,[16-23] where time-multiplexed channel switching has been employed for sequential data acquisition, the transfer curves of all devices in our setup are acquired simultaneously across channels, without temporal offsets between measurements. Measurements were performed using a 16-channel SMU system, with one channel assigned to the electrolyte-gate voltage and 15 channels used for independent source–drain biasing and current readout of the GFET devices (see Methods for details). A comparison of the reported readout strategy with the literature data is presented in Supporting Information Section 8 and Table S4. Note that the gate-source leakage current ($I_{sg}$) in our devices was measured to be below ±60 nA (see Figure S2 in the Supporting Information), and thus had a negligible influence on the reported source-drain currents, which were in the ~10 µA range.

*Real-time Measurement of Antibody Immobilization*

Next, we present real-time monitoring of the immobilization of SARS-CoV-2 S-protein antibody, $Ab_{(S)}$, on the $N_3$-CNM/GFET surface (Figure 2). To this end, some of the devices on the same chip were functionalized with DBCO-labeled C-reactive protein antibodies ($Ab_{(Ref)}$), which are similar in size and weight to the $Ab_{(S)}$ (see Figure S3 in the Supporting Information). These $Ab_{(Ref)}$-functionalized GFETs act as reference devices, enabling measurement of the differential sensor response mitigating parasitic effects, including non-specific adsorption of the biomolecules.

Figure 2a shows the time-dependent $\Delta V_{CNP}(t)$ values ($\Delta V_{CNP}(t) = V_{CNP}(t) - V_{CNP}(t=0)$) for signal (blue symbols) and reference (red symbols) GFETs measured simultaneously during the injection of $Ab_{(S)}$. Prior to the injection of $Ab_{(S)}$, the transfer curves of both signal and reference devices were recorded in parallel in the measurement buffer (3000-fold diluted phosphate-buffered saline, 0.0003X PBS-P, pH 7.4, 20 µL/min). After completing each forward and backward scan cycle (gate sweep step size of 2 mV), both curves were fitted in real time using the measurement software (see Supporting Information, Section 7 for details) to extract the charge neutrality point ($V_{CNP}$), thereby generating real-time $V_{CNP}$ sensorgrams for all the devices. Upon injection of $Ab_{(S)}$ solution (100 µg/mL in 10 mM sodium acetate, pH 4.5, 5 µL/min), all devices exhibited a pronounced negative shift in $V_{CNP}$. This shift arises from the buffer exchange (PBS-P to sodium acetate). The devices were maintained in $Ab_{(S)}$ solution for 30 min before the PBS-P buffer was reintroduced. After the buffer exchange to PBS-P, the solution bulk effect was compensated, and a progressive separation in $V_{CNP}$ between signal and reference devices was observed, indicating specific antibody binding on the signal devices. The corresponding real-time differential sensorgram (signal–reference) is shown in Figure 2b, where a relative downshift of ~25.8 mV is observed between the two stable buffer baselines (t = 0 min and t = 65 min). The increase in $\Delta V_{CNP}$ during $Ab_{(S)}$ binding in acetate (t = 10–35 min), followed by the decrease after switching to PBS-P, can be attributed to the changes in the antibody's effective interfacial charge. These changes arise from pH-dependent protonation and differences in ionic strength between the two buffers, which modulate the fraction of antibody charge that couples to the graphene channel.[43] Representative transfer curves of a signal device, with $V_{CNP}$ values referenced against a control device, are displayed in Figure 2c. This shift is attributed to antibody binding on the signal devices, whereas the majority of $N_3$ sites on the reference devices were already blocked by the $Ab_{(Ref)}$ antibodies. The specific $Ab_{(S)}$ immobilization on the $N_3$-CNM/graphene surface was further evaluated under physiological PBS-P buffer conditions, revealing a pronounced $V_{CNP}$ downshift compared to bare graphene reference devices. Detailed results and supporting data are provided in Figure S4 in the Supporting Information.

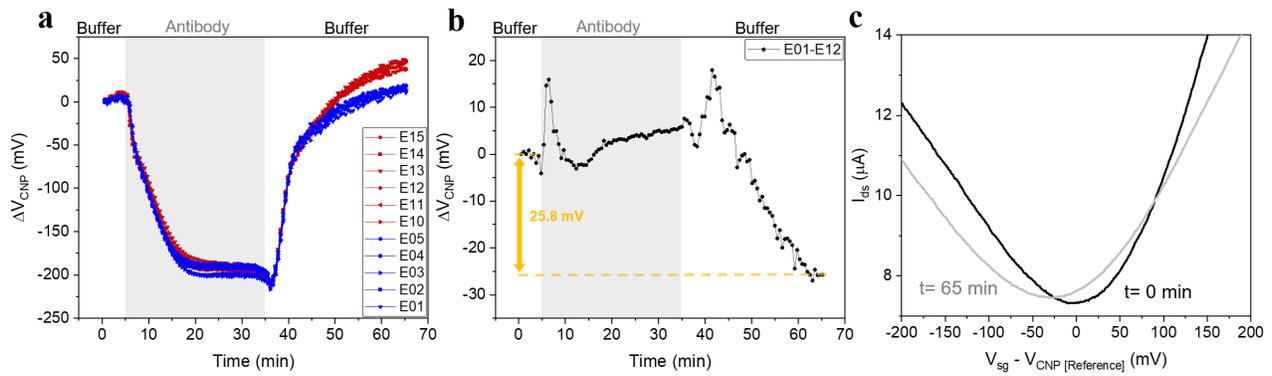

**Figure 2**. Real-time monitoring of SARS-CoV-2 S-protein antibody immobilization on $N_3$-CNM/graphene FETs. a) Time dependent $\Delta V_{CNP}(t)$ sensorgrams ($\Delta V_{CNP}(t) = V_{CNP}(t) - V_{CNP}(t=0)$) measured in parallel for signal devices (blue symbols) and reference devices (red symbols). The sections labeled "buffer" correspond to measurements in diluted PBS-P buffer (0.0003X PBS-P), before and after the antibody injection (highlighted in grey) b) Real-time differential sensorgram ($V_{CNP\,[Signal]} - V_{CNP\,[Reference]}$) for a typical pair of signal-reference devices in a, where a relative sensorgram shift of 25.8 mV in the measurement buffer marked by the yellow arrow. c) Corresponding transfer curves ($V_{ds}$= 10 mV) at t = 0 min (black curve) and t = 65 min (grey curve) for the signal device used to calculate the response in b, with $V_{CNP}$ values shifted relative to a reference device.

*Real-Time Antigen Detection*

Figure 3a illustrates the functionalization of an $N_3$-CNM/GFET chip by drop-casting Ab(s) and $Ab_{(Ref)}$ solutions (both with 100 µg/mL concentration in 10 mM sodium acetate buffer, pH 4.5), which serve as signal and reference devices, respectively. To ensure that this procedure does not compromise antigen recognition, we verified that the binding capability of antibodies on $N_3$-CNM/graphene is preserved throughout the GFET fabrication process, including CNM transfer and subsequent antibody immobilization. Surface plasmon resonance (SPR) measurements confirmed similar dissociation constants ($K_D$) of N-protein and S-protein for two fabrication strategies: (1) direct synthesis of $N_3$-CNM on gold SPR slide followed by *in situ* antibody functionalization, and (2) transfer of $N_3$-CNM/graphene heterostructures on gold SPR slide followed by antibody drop-casting (*ex situ*). Detailed results are provided in the Supporting Information, Section 3 and Figure S5.

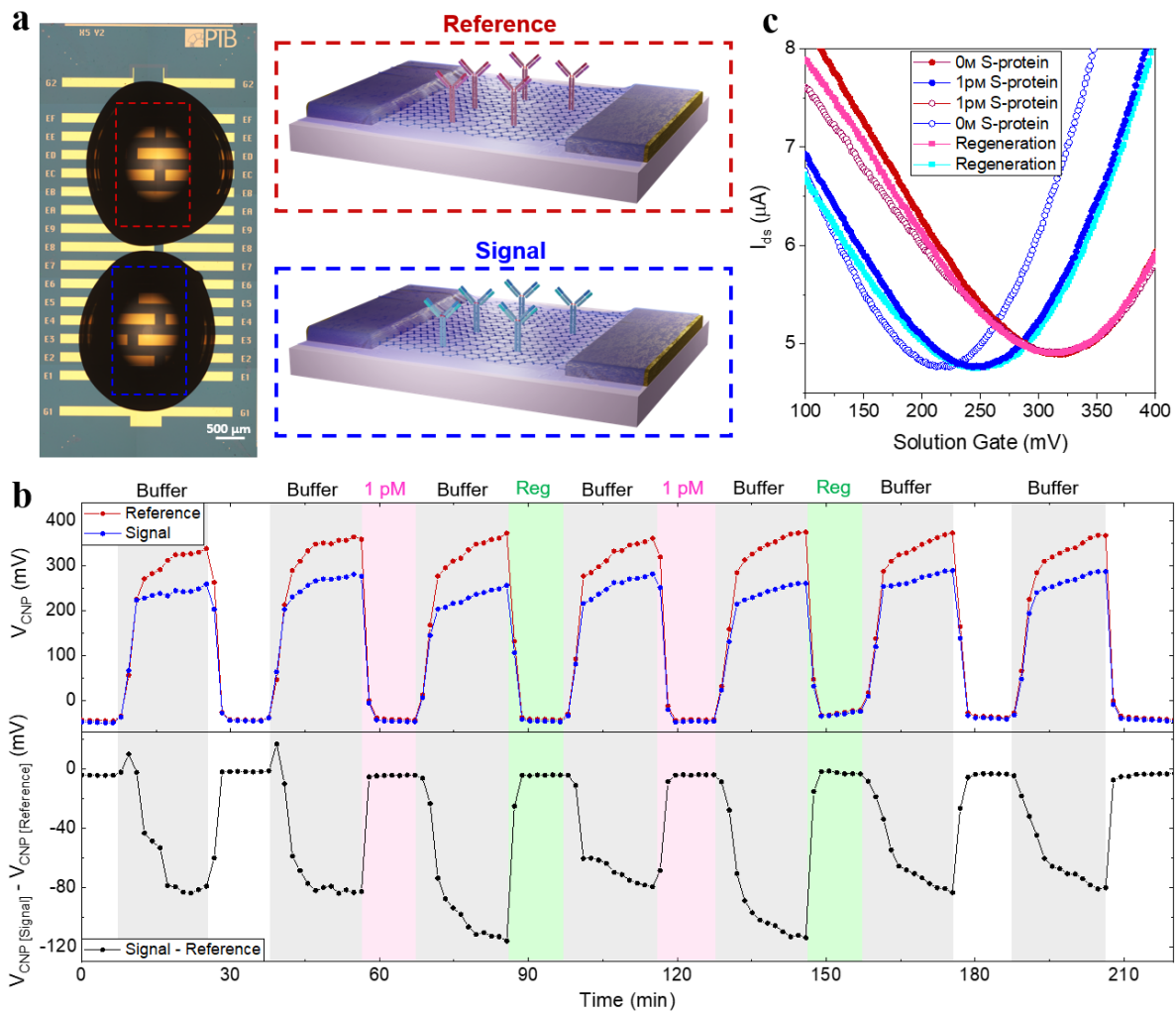

**Figure 3**. Real-time S-protein detection using antibody-functionalized $N_3$-CNM/GFETs. a) Optical image and schematic representation of a GFET chip pre-functionalized with $Ab_{(Ref)}$ (reference devices) and $Ab_{(S)}$ (signal devices). b) The upper panel displays $V_{CNP}$ sensorgrams for representative signal (in blue) and reference (in red) devices measured simultaneously, while the lower panel shows their corresponding differential sensorgram. The grey shadings mark the injection periods of the measurement buffer (0.0003X PBS-P), while pink and green shades highlight the injection periods for S-protein solution (1 pM in 1X PBS-P) and the regeneration process, respectively. c) Shift in the transfer curves of representative $Ab_{(Ref)}$- and $Ab_{(S)}$-functionalized devices measured in 0.0003X PBS-P before and after the injection of 1 pM S-protein and after regeneration.

Figure 3b (upper panel) displays sensorgrams, $V_{CNP}(t)$, for two devices: $Ab_{(S)}$-functionalized (signal, blue symbols) and $Ab_{(Ref)}$-functionalized (reference, red symbols). The sensorgrams begins with a 10-minute measurement in physiological buffer (1X PBS-P, 60 µl/min), followed by a 20-minute measurement in diluted buffer (0.0003X PBS-P, 60 µl/min). This transition from high to low ionic strength causes a remarkable shift in the $V_{CNP}$ of both devices towards higher gate voltages due to the decrease of the double layer capacitance, which leads to a weaker capacitive coupling between the graphene layer and the gate electrode. To enable both biorecognition and highly sensitive measurement, we adopt an optimal approach that involves maintaining physiological conditions during the binding process and sensing measuring in diluted buffers. This cycling protocol aims to mitigate the charge screening effect while preserving the biomolecule's functionality and structural integrity.

In the physiological buffer a difference in the doping level between the reference and signal devices is negligible, since Debye screening effectively masks it. However, in the diluted buffer this difference becomes distinct, where the $V_{CNP}$ of the signal device is notably lower than that of the reference, which agrees with the results of real-time antibody immobilization measurement shown in Figure 2a. The observed doping difference leads to a negative shift in the differential sensorgram (Figure 3b, lower panel). To assess the stability of the sensor baseline, the injection of buffer was repeated in several cycles before introducing the target antigen. Both reference and signal devices show a slight $V_{CNP}$ increase during the second cycle compared to the first. However, their differential response remains consistent with the first cycle. These results highlight the necessity of reference correction when operating in low-ionic-strength buffers. Since signal and reference devices are measured simultaneously, they experience a similar $V_{CNP}$ drift thus their subtraction yields a stable baseline in the differential sensorgram. Hysteresis effects observed at low ionic strength[44-47] and the corresponding correction approach are discussed in the Supporting Information, Section 5.

Next, the target antigen (1 pM S-protein concentration in 1X PBS-P, 60 μl/min) was introduced under the physiological conditions for optimal binding. During this phase, the change in the differential sensorgram is minimal. However, when the binding is measured in the diluted buffer, a shift towards lower voltage in the differential sensorgram is observed. This is due to the positive charge of the S-protein,[48] leading to an increased electron-doping in the signal GFET, thereby causing the $V_{CNP}$ to shift to lower gate voltages. In the next injection cycle, the sensor surface is regenerated by dissociating the bound antigens using a 10 mM Glycine–HCl solution (see Supporting Information, Section 1 for details). Following the surface regeneration (indicated as Reg in Figure 3b), the differential sensorgram recorded in the diluted buffer exhibits a shift towards higher voltages as expected.

In the next steps, the cycles of antigen injection and surface regeneration were repeated, consistently showing similar binding and recovery responses. The detection measurements with the SPR technique demonstrated that the antibody maintains its antigen-binding functionality for up to five regeneration cycles.[31] In the final cycle, following the injection of the 0 M target concentration (blank 1X PBS-P), the response in the differential sensorgram remains stable, further indicating the measurement's reliability in the diluted buffer environment. The transfer curves of representative reference (red) and the signal (blue) devices measured concurrently in the diluted buffer are shown in Figure 3c. The $V_{CNP}$ downshift of the signal device after the antigen introduction (1 pM S-protein) is restored to its initial state (0 M S-protein), while the $V_{CNP}$ of the reference device is nearly unchanged.

*Detection of Concentration Series of S- and N-protein*

Using the presented detection methodology, we conducted real-time measurements of various antigen concentrations (in 1X PBS-P, 60 µl/min) and diluted buffer (0.0003X PBS-P, 60 µl/min) through successive injection cycles. Figure 4a (upper panel) displays the $V_{CNP}$ sensorgram for devices pre-functionalized with Ab$_{(S)}$ (complementary antibody, four signal devices, blue symbols) and Ab$_{(Ref)}$ (non-complementary antibody, four reference devices, red symbols). The lower panel shows the differential sensorgram for a typical signal-reference pair. The sensor's zero-baseline is established before antigen injections by measuring $V_{CNP}$ shifts when transitioning from the blank 1X PBS-P (0 M S-protein concentration) to the diluted buffer. Subsequently, the sensor response to S-protein concentrations ranging from 1 aM to 100 pM, increasing by an order of magnitude at each step, is measured. In the diluted buffer, a distinguishable downshift response in the differential sensorgram becomes apparent after the introduction of 10 aM S-protein solution. This response further increases up to 1 pM S-protein solution and remains nearly constant for higher concentrations, indicating the saturation. The decrease in hole-doping of the signal GFETs with increasing the target concentrations corresponds with the net positive charge of the S-protein[48] (see Figure S6 in the Supporting Information for changes of a typical signal device transfer curve).

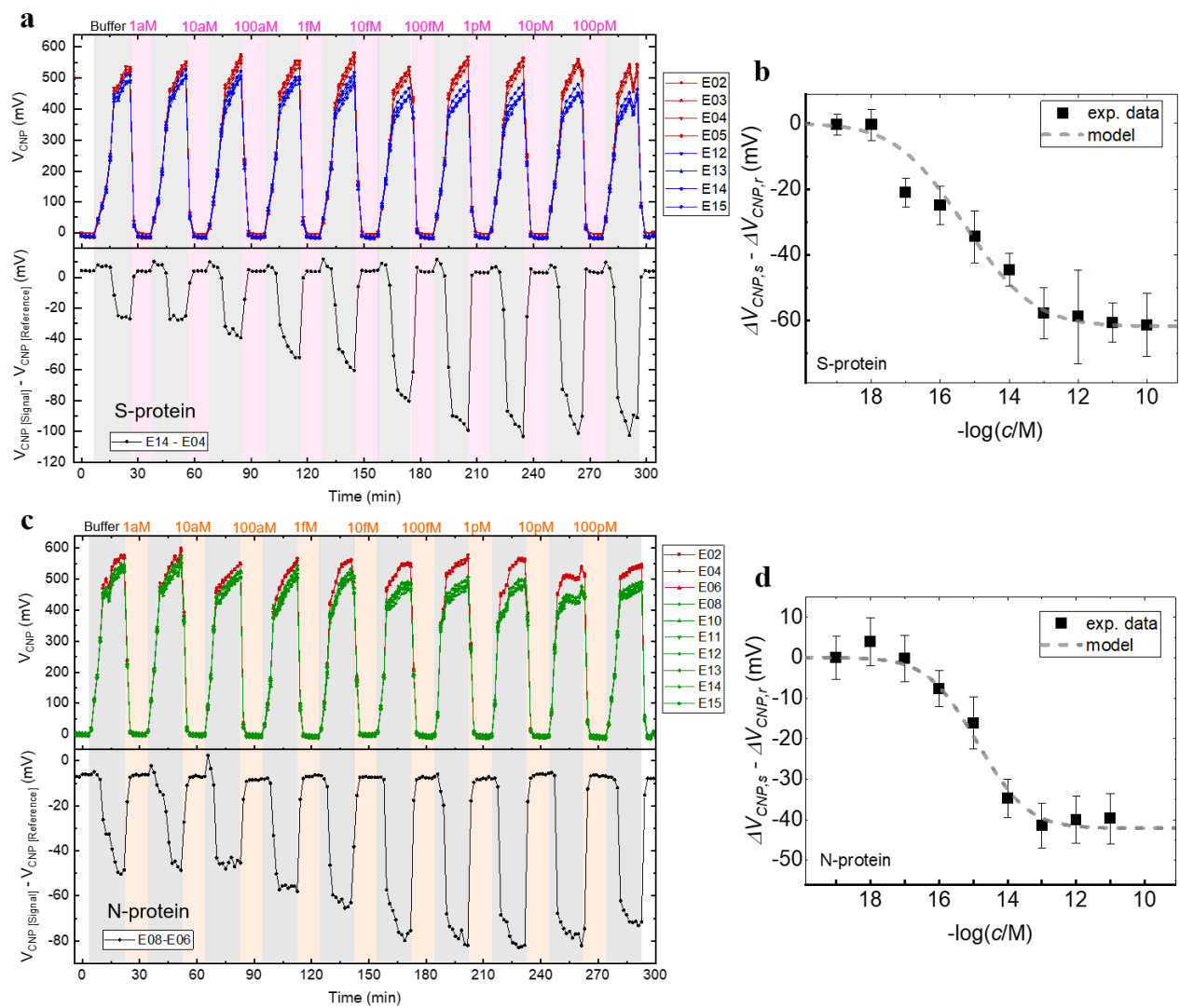

**Figure 4.** Concentration series analysis of S- and N-protein detection. a) Upper panel: $V_{CNP}$ sensorgrams for series of S-protein concentrations with four signal devices in blue and four reference devices in red; lower panel: differential sensorgram of a typical signal-reference pair. c) Upper panel: $V_{CNP}$ sensorgrams for series of N-protein concentrations with signal devices in green and reference devices in red; lower panel: differential sensorgram of a typical signal-reference pair. Grey shadings indicate measurement buffer injection periods; pink and orange shadings in (a) and (c) highlight S- and N-protein injection periods, respectively. The average response to different concentrations of b) S-protein and d) N-protein. The grey dashed line indicates the model fitting results.

The average differential response obtained from all 16 signal–reference combinations (4 signal devices × 4 reference devices) for concentration-dependent S-protein detection is shown in Figure 4b, with the response to the blank buffer normalized to zero. The $V_{CNP}$ for the sample with 10 aM S-protein shifts by $V_{CNP}(10aM) - V_{CNP}(0M) = -20.6 \pm 4.4\ mV$ to lower voltages as compared to the sample without S-protein. By increasing the concentration to 100 aM, 1 fM, 10 fM and 100 fM, the transfer curves shift further by $V_{CNP}(100aM) - V_{CNP}(10aM) = -3.9 \pm 5.8\ mV$, $V_{CNP}(1fM) - V_{CNP}(100aM) = -9.7 \pm 8.0\ mV$, $V_{CNP}(10fM) - V_{CNP}(1fM) = -10.0 \pm 5.0\ mV$ and $V_{CNP}(100fM) - V_{CNP}(10fM) = -13.3 \pm 7.7\ mV$, respectively. By applying the concentration of 1 pM, 10 pM, and 100 pM, the $V_{CNP}$ shift in comparison to the 100 fM sample is only $-1.0 \pm 14.3\ mV$, $-2.8 \pm 6.0\ mV$, and $-3.5 \pm 9.7\ mV$, which demonstrates saturation of the binding places for the S-protein on the surface. Thus, the sensor shows a dynamic range of 4 orders of magnitude and a limit of detection, defined as the target concentration where the response reaches 1X the standard deviation, is approaching 10 aM concentration of S-protein.

Concentration-dependent measurements were performed analogously for N-protein detection. Figure 4c shows the corresponding $V_{CNP}$ sensorgrams, comparing signal devices functionalized with N-protein antibodies (Ab$_{(N)}$, seven signal devices, green symbols) to reference devices (Ab$_{(Ref)}$, three reference devices, red symbols). Similar to the S-protein case, Ab$_{(N)}$-functionalized devices exhibit a baseline shift toward lower gate voltages in diluted buffer, reflecting differences in initial doping levels. In the differential sensorgrams, a clear downshift is observed upon exposure to N-protein, with the first detectable response appearing at 100 aM. Figure 4d summarizes the averaged differential response obtained from 21 signal–reference combinations, normalized to the blank buffer. At 100 aM N-protein, the $V_{CNP}$ shifts by $-7.6 \pm 4.2\ mV$ relative to the blank. Further increases in concentration to 1 fM, 10 fM, and 100 fM result in additional shifts of $-8.5 \pm 6.4\ mV$, $-18.7 \pm 4.7\ mV$, and $-6.7 \pm 5.5\ mV$, respectively. At higher concentrations (1 pM and 10 pM), only marginal changes are observed, indicating saturation of the available N-protein binding sites. Overall, the sensor exhibits a dynamic range of approximately 2–3 orders of magnitude for N-protein detection, with a limit of detection approaching 100 aM.

*Model Analysis of the Sensor Response*

To quantitatively analyze the GFET sensor response, we applied a model, which describes the GFET/electrolyte interface as a system of three capacitively coupled layers – the graphene channel, the CNM/antibody layer, and the electrical double layer – which together must satisfy overall charge neutrality.[27] In the model the binding of targets to the antibodies is described using a Langmuir–Freundlich adsorption isotherm, which can account for heterogeneous binding affinity distributions. To accurately account for the electrostatic interactions between the target and the sensor surface, we determine and explicitly consider the charge of the antibodies, see the Supporting Information Section 4. Using the equilibrium dissociation constants $K_D$ obtained from SPR measurements as input, we calculate the sensor response, which is found in a good agreement with the experimental data, see Figure 4b,d. All model parameters, summarized in Table S3, are found in qualitative agreement with the expected properties. The extracted antibody charge is negative, consistent with the typically negative charge of IgG antibodies at pH 7.4, see Ref. [49]. The extracted positive target charge of the N-protein agrees qualitatively with its net charge at pH 7.4, see the Supporting Information Section 6, and the extracted positive charge for the S-protein is in a good agreement with literature values.[48] Moreover, based on the model we estimate the limits of detection of ~5 aM and ~95 aM for the S-protein and for the N-protein, respectively.

*Toward multiplexing: Selective detection of S- and N-proteins with a Single GFET chip*

The simultaneous detection of SARS-CoV-2 S-protein and N-protein using multiplex methods provides valuable insights, particularly for viral mechanism studies, the study of long COVID and vaccine development.[50, 51] Building on this need for multiplexed analysis, our multi-device GFET architecture, combined with parallel device-specific readout, enables selective detection of both S- and N-proteins using a single GFET chip. Figure 5a (upper panel) displays the $V_{CNP}$ sensorgrams for GFETs functionalized with three different types of antibodies. Certain devices are pre-functionalized with $Ab_{(S)}$ (devices E12-E14, blue symbols) and others with $Ab_{(N)}$ (devices E07-E09, green symbols) as complementary antibodies for specific detection of the corresponding proteins. Additionally, a device is pre-functionalized with $Ab_{(Ref)}$ (device E03, red symbol) serving as non-complementary reference.

Initially, the sensor's zero-baseline is established in a blank diluted buffer (0.0003X PBS-P, 60 µl/min). Following this, 1 pM S-protein and then 1 pM N-protein (in 1X PBS-P, 60 µl/min) are injected sequentially, with their binding responses measured in the diluted buffer. The differential sensorgrams for devices functionalized with $Ab_{(S)}$ and $Ab_{(N)}$ during these injections are illustrated in the middle and lower panels of Figure 5a, respectively. For the $Ab_{(S)}$ functionalized device, a downshift response is observed post S-protein injection, remaining largely unchanged after N-protein injection. Conversely, for the $Ab_{(N)}$ functionalized devices, the downshift response is only evident following the injection of N-protein. Figure 5b displays typical transfer curves (normalized to the reference device's $V_{CNP}$) measured in the diluted buffer for devices functionalized with $Ab_{(S)}$ and $Ab_{(N)}$ in its upper and lower panels, respectively.

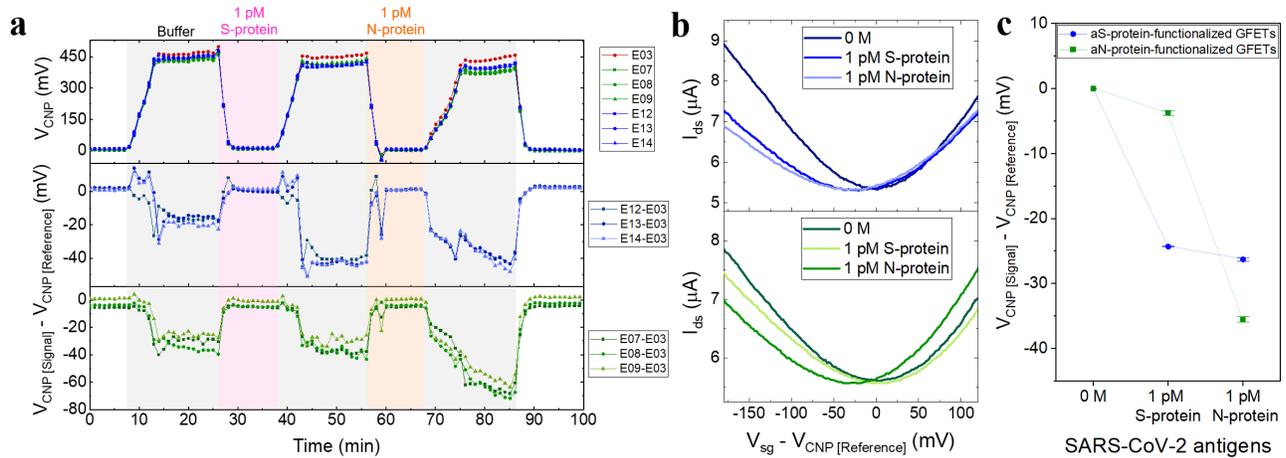

**Figure 5.** Dual detection of SARS-CoV-2 S- and N-protein using a single GFET chip. a) (upper panel) Real-time $V_{CNP}$ sensorgrams for a reference device functionalized with Ab$_{(Ref)}$ (E03, red symbol), three Ab$_{(N)}$-functionalized devices (E07-E09, green symbols) and three Ab$_{(S)}$-functionalized devices (E12-E14, blue symbols). The grey shadings mark the injection periods of the measurement buffer (0.0003X PBS-P), while pink and orange shades highlight the injection periods for S-protein and N-protein solutions (1 pM in 1X PBS-P), respectively. The middle and lower panels represent the differential sensorgrams for devices functionalized with Ab$_{(S)}$ (blue symbols) and Ab$_{(N)}$ (green symbols), respectively. b) Typical transfer curves of Ab$_{(S)}$ (upper panel) and Ab$_{(N)}$ (lower panel) functionalized devices obtained in 0.0003X PBS-P buffer after sequential injection of 0 M, 1 pM S-protein and 1 pM N-protein in 1X PBS-P. c) Average $\Delta V_{CNP}$ response for Ab$_{(S)}$ (blue symbol) and Ab$_{(N)}$ (green symbol) functionalized devices after the sequential injections.

Using the method outlined in the previous section for determining the reporting points in differential sensorgrams, Figure 5c plots the average response of three signal-reference pairs for each protein. The sensor, on average, demonstrates a response of 24.3 mV for 1 pM S-protein in the complementary device and a 3.8 mV cross-reactivity response in the non-complementary $Ab_{(N)}$ functionalized device. For N-protein, the response in the complementary device is 31.7 mV while 2.1 mV cross-reactivity is measured in the $Ab_{(S)}$ functionalized device. These results indicate distinctly detectable responses with low cross-reactivities in dual detection of SARS-CoV-2 S- and N-protein using a single GFET sensor.

In addition, we engineered a compact, integrated portable platform for point-of-care testing (see Figure S11a in the Supporting Information), where electronic components and microfluidic systems are unified into a single unit. To validate this setup, we conducted a proof-of-concept test, detecting 1 pM N-protein (in 1X PBS-P, 60 µl/min) using 0.0003X PBS-P as the measurement buffer. Before antigen injection, a noticeable doping difference ($\Delta V_{CNP}$) is evident between the signal devices (functionalized with $Ab_{(N)}$) and the reference devices (functionalized with $Ab_{(Ref)}$), as illustrated in Figure S11b in the Supporting Information. This difference notably increases by an average of 23.2 mV following antigen binding (see Figure S11c in the Supporting Information). These results align well with findings from previous experiments, supporting the detection using the portable platform.

**Conclusions**

In summary, we developed a scalable sensing method for real-time and ultra-sensitive detection of SARS-CoV-2 proteins using CNM-GFET heterostructures. This platform enables site-specific covalent immobilization of antibodies on carbon nanomembranes while preserving the intrinsic electronic properties of the underlying graphene. The integration of an automated microfluidic system with a parallel electrical readout allows simultaneous monitoring of multiple GFETs on a single chip with high temporal resolution and fidelity required for robust kinetic and statistical analyses. Measurements employing built-in reference channels were used to compensate for bulk effects, baseline drift, and non-specific signals, enabling reliable differential sensorgram analysis. By quantitatively analyzing signal changes between two different buffer solutions, we estimate the electrical charge density of the antibodies and targets and calculated the specific binding response using an analytical model. The sensor achieved limits of detection of 10 attomolar for the S-protein and 100 attomolar for the N-protein, and demonstrated their dual detection with negligible cross-reactivity. Based on the developed biofunctionalization strategy and electrical readout method, we further demonstrated a proof-of-concept portable sensing setup suitable for point-of-care testing. The combination of CNM-based biofunctionalization, parallel real-time GFET readout, and automated microfluidics establishes a strong foundation for future multiplexed diagnostics and portable point-of-care applications.

## Methods

**Fabrication of N$_3$-CNM/GFET Van der Waals Heterostructure Devices**

The microfabrication process for GFET chips is detailed in Ref.[26] Briefly, the monolayer graphene sheet (Graphenea) on the SiO$_2$/Si wafers was patterned using standard electron beam lithography (EBL, Vistec EBPG 5000+) with a polymethyl methacrylate (PMMA) resist (AR-P 671.04, Allresist). The pattern transfer was performed by dry etching in an argon/oxygen plasma. To directly contact the graphene channel and hence lower the contact resistance, 80/20 nm Au contacts were thermally evaporated in a two-step lithographic process while employing 2 nm of Ti as an adhesion promoter.

To synthesize the azide-terminated CNMs (N$_3$-CNM), first 4'-nitro-[1,1']-biphenyl-4-thiol (NBPT) monolayers were self-assembled on gold substrates. These were then exposed to low-energy electron irradiation in a high vacuum (<5×10$^{-7}$ mbar), transforming the nitro groups into amino groups, and leading to the formation of NH$_2$-CNMs.[52] Subsequently, these CNMs were functionalized by attaching azidoacetyl chloride (2-AAC, 97%, 30% solution in diethyl ether, SelectLab Chemicals GmbH, Münster, Germany) to the amino terminals as detailed in Ref.[27] In brief, the transfer of N$_3$-CNM was achieved using a wet chemical etching method. Two layers of PMMA were spin-coated onto the N$_3$-CNM. The first layer used PMMA with a low molecular weight of 50K, and the second layer used a high molecular weight PMMA of 950K. Both layers were applied for 1 minute at 2000 rpm. After spin-coating, the N$_3$-CNM covered with PMMA was air-dried, considering its sensitivity to higher temperatures. Subsequently, the mica support, which underlays the gold/N$_3$-CNM/PMMA structure, was separated by immersing the assembly in water. The gold substrate was then dissolved using an etching bath of I$_2$/KI/H$_2$O in a 1:4:10 concentration ratio, a process that took approximately 15 minutes. Following this, the N$_3$-CNM was cleaned in pure water and carefully transferred onto a GFET chip. To ensure robust adhesion to the substrate, the $N_3$-CNM was left to dry in ambient condition overnight. The final step involved dissolving the PMMA layer by soaking it in acetone for one hour followed by rinsing with isopropyl alcohol (IPA).

**Measurement Electronics**

The measurement electronics for the GFET arrays incorporates a total of 31 channels to interface with 15 GFETs, each having two contacts for drain and source, and one connection for the gate electrode. The measurement unit precisely adjusts and measures both positive and negative voltages within the single-digit volt range, achieving a precision of at least 500 µV and a current accuracy of 1 nA. The core component of the system is an integrated circuit (IC) designed specifically for parametric measurement units (PMU) (see Figure S12 in the Supporting Information). This IC provides voltage or current on four channels simultaneously and measures the complementary quantity using an analog-to-digital converter (ADC). The measurement unit consists of a circuit board equipped with the PMU chip and a four-channel ADC chip. To achieve the required 31 channels, eight circuit boards were mounted on a custom-designed carrier board. Electrical connections to the GFETs were established *via* cables. Detailed procedures for the measurement program and the readout architecture are provided in the Supporting Information Section 7 and 8. Liquid injections were regulated *via* an automated microfluidic system (Elveflow®), equipped with a flow controller (OB1), 12 reservoirs for samples and buffer solutions, a distributor valve (MUX) and a flow sensor with 0-80 µl/min sensitivity range.

**Supporting Information**

Material, Preparation of N$_3$-CNM/graphene/gold SPR sensor slide, SPR measurements, Charge transfer model for the GFET response, Hysteresis suppression through differential readout in GFET arrays, Calculation of the charge of the N-protein, Data acquisition for the GFET sensor response, Comparison of readout architecture for GFET arrays

**Author Contributions**

A.T. devised the project and experimental plan. H.R.R. and D.K. performed the GFET measurements, analyzed the data and interpreted results. D.K. developed model calculations. G.E. performed chemical functionalization and SPR measurements. M.M., M.R. and A.R. developed the measurement electronics and software. K.F. and D.G. prepared biological samples for functionalization and testing. C.N. performed AFM measurements and analysis. The manuscript was written by HRR, DK and AT with contributions and comments from all co-authors.

**Notes**

The authors declare no competing financial interest.

**Acknowledgments**

We acknowledge the financial support of this work by BMWK project 2163 BR / FE 2 "ViroGraph", BMBF project 13N15744 "SARS-CoV-2Dx" as well as by ESF project 2024 FGR 0069 "Multi-Interact" and the European Funds for Regional Development (EFRE 2021-2027; Project FGI 0010, 2DMat-Lith-Lab). We thank Prof. Ralf Ehricht from Friedrich-Schiller-University Jena for help with the identification of the amino acid sequence of the SARS-CoV-2 N-protein.

# Supporting Information

# Ultrasensitive Real-Time Detection of SARS-CoV-2 Proteins with Arrays of Biofunctionalized Graphene Field-Effect Transistors

*Hamid Reza Rasouli[1], David Kaiser[1], Ghazaleh Eshaghi[1], Marco Reinhard[2], Alexander Rolapp[2], Dominik Gary[3], Tobias Fischer[3], Christof Neumann[1], Thomas Weimann[4], Katrin Frankenfeld[3], Michael Meister[2], Andrey Turchanin[1,5]\**

[1]Institute of Physical Chemistry, Friedrich Schiller University Jena, 07743 Jena, Germany

[2]IMMS Institut für Mikroelektronik- und Mechatronik-Systeme gemeinnützige GmbH (IMMS GmbH), 99099 Erfurt, Germany

[3]fzmb GmbH, Forschungszentrum für Medizintechnik und Biotechnologie, 99947 Bad Langensalza, Germany

[4]Physikalisch-Technische Bundesanstalt (PTB), 38116 Braunschweig, Germany

[5]Jena Center for Soft Matter (JCSM), 07743 Jena, Germany

**Corresponding Author:** Andrey Turchanin (andrey.turchanin@uni-jena.de)

**Materials and Methods**

1. Materials

4'-Nitro-[1,1']-biphenyl-4-thiol (NBPT) was procured from Taros Chemicals GmbH. N,N-Dimethylformamide, with a purity of 99.8% and designated as extra dry AcroSeal™, was obtained from VWR International GMBH. Azidoacetyl chloride, with a 30% solution in Ether, was obtained from SelectLab Chemicals GmbH. N,N-Diisopropylethylamine, subjected to redistillation, was acquired from Sigma-Aldrich Chemie GmbH. Tetrahydrofuran (99.5%, extra dry over molecular sieve, stabilized, AcroSeal™) was purchased from Thermo Fisher Scientific. DBCO-labeled monoclonal antibodies targeting the nucleocapsid protein ($Ab_{(N)}$), spike protein ($Ab_{(S)}$), human-C-reactive protein ($Ab_{(Ref)}$) and receptor-binding domain of S-protein were obtained from Forschungszentrum für Medizintechnik und Biotechnologie (fzmb GmbH). The SARS-CoV-2 N-protein was obtained from Virion/Serion GmbH. Phosphate-buffered saline (PBS-P + Buffer 10x), Glycine/HCl at 10 mM with a pH of 2, and Immobilization buffer (10 mM Sodium acetate at pH 4.5) were purchased from Cytiva GmbH. The use of 10 mM sodium acetate buffer at pH 4.5 for antibody immobilization follows established protocols for covalent coupling to carboxylated surfaces, where proteins are pre-concentrated in the matrix by electrostatic attraction when dissolved in low-ionic-strength buffer at a pH below their isoelectric point (pI). At pH 4.5, which lies below the pI of IgG antibodies, they carry a net positive charge and are attracted to the negatively charged surface functional groups, thereby enhancing immobilization efficiency.[1]

Serial dilutions of the antigen solutions were prepared in 1X PBS-P buffer using a standard 10-fold dilution protocol to ensure accurate and reproducible concentrations for all experiments. Beginning from a well-defined stock solution, each subsequent dilution was produced by mixing one part of the preceding solution with nine parts of fresh buffer while using low-binding plasticware, calibrated micropipettes, and freshly prepared aliquots to minimize adsorption and dilution errors. This stepwise approach allows precise preparation of extremely low concentrations, including those in the attomolar regime, where all dilutions were performed immediately before use to ensure stability and reliability of the resulting solutions. The same rigorously controlled procedure was applied for every concentration series used throughout this study.

Surface regeneration was performed using a 10 mM Glycine–HCl solution. Prior to regeneration, the system was equilibrated in 1X PBS-P under physiological conditions. The Glycine–HCl solution was then applied for 30 seconds to dissociate the antigen–antibody complexes, followed by reintroduction of 1X PBS-P. This procedure promotes conformational recovery of the immobilized antibodies in a high-ionic-strength environment after antigen removal.

## 2. Preparation of $N_3$-CNM/graphene/gold SPR sensor slide

CVD-graphene was transferred using the electrochemical delamination method to the target substrates (gold SPR sensor slide). The transfer was performed using poly (methyl 2-methylpropenoate) (PMMA) (AR-P 671.04) dissolved in chlorobenzene as a supporting layer.hamid[2] The PMMA with a molecular weight of 450 K was spin-coated for 1 min at 2000 rpm and then dried on a hot plate at 90 °C for 15 min. Graphene was detached from the Cu substrate by applying a voltage of 2-3 V in 200 mM NaOH, as described in Ref.[3]. This transfer was used for samples with an area of 1×1 cm$^2$. After the removal of the growth substrate, the PMMA/graphene stack was washed three times in degassed water for 15 minutes and fished out from the solution with the gold SPR sensor slide. Next, the sample was dried on a hotplate at 180 ºC for 3 hours to improve the adhesion with the substrate. The PMMA was removed by dipping the sample in acetone for 30 minutes followed by rinsing with isopropyl alcohol (IPA). Finally, the $N_3$-CNM was transferred onto a graphene/gold substrate, following the procedure outlined in Methods section.

## 3. SPR measurements

A multiparametric SPR instrument (Navi 210 A BioNavis) was employed to observe the interaction between the recognition elements and the target molecules. All evaluations were conducted by performing continuous full-range angular scans (40−78°) using the laser wavelength 670 nm. For the experiments shown in Figures S7a and S7c, $N_3$-CNM was synthesized on the Au sensor slide without any transfer. For the experiments shown in Figures S7b and S7d, first graphene was transferred on the Au sensor slide followed by transfer of $N_3$-CNM. For the experiments shown in Figures S7a and S7c, the SPR sensor slide was then further functionalized with 100 µg/ml antibody solution in 10 mM Sodium acetate (pH 4.5) in the signal channel followed by Casein in the signal and reference channel. For the experiments shown in Figures S7b and S7d, the functionalization with the antibody was done by drop casting *ex situ* for the signal channel flow cell. The N-protein and S-protein concentration series were injected in physiological buffer at a speed of 10 µl/min for 5 and 15 minutes with 5 and 10 minutes dissociation time, respectively. After injection of each S-protein concentration regeneration buffer (Glycine/HCl at 10 mM with a pH of 2) was injected for 30 seconds at 10 µl/min. The temperature was kept stable at 25 °C. The response at the equilibrium states of the SPR angle shifts ($\Delta\theta_{SPR}$) was recorded. Figure S5 presents the reference corrected binding response $\Delta\theta_{SPR[Signal]}$ - $\Delta\theta_{SPR[Reference]}$ of the N- and S-protein to their complementary antibody on $N_3$-CNM and $N_3$-CNM/graphene surfaces. The curve fitting was done with TraceDrawer$^{TM}$ 1.10 using the Langmuir 1:1 binding model.

In our previous study, we synthesized the $N_3$-CNM directly on the Au sensor slide and performed antibody functionalization in-line.[4] Here, we have included the transfer of graphene onto the Au sensor slide, the transfer of $N_3$-CNM onto the graphene to form the graphene/$N_3$-CNM van der Waals heterostructure, and *ex situ* antibody functionalization by dropcasting technique.

Consequently, the antibody-functionalized surface used in the SPR validation experiments closely resembles the GFET surface. The primary differences are that the substrate for SPR experiments is Au, whereas $SiO_2$ is used in the GFET experiments, and no electrical potential is applied to graphene or the solution in the SPR measurements. Across various experiments, we observed no systematic difference between the binding constants calculated for the experiments with and without graphene below the $N_3$-CNM see Figure S5 and Table S1. This aligns with previous studies indicating that the presence of graphene enhances the binding response only in the phase of the SPR sensorgram,[5] but not in the shift of the minimum angle, which is the focus of our analysis. Note that the $Ab_{(S)}$ used for the SPR measurements shown in Figures S7c and S7d differs from the one utilized in the FET experiments. This was necessitated by the unavailability of the original antibody at the time of these specific SPR validation experiments. The FET measurements, however, were conducted with the same $Ab_{(S)}$ as described in our previous work.[4] Despite this difference, the results show the negligible influence of transfer, functionalization methods (*in situ*, *ex situ*), and the presence of graphene underneath the $N_3$-CNM layer.

**Table S1.** Kinetic rates for the association, $k_a$, and dissociation, $k_d$, as well as dissociation constants, $K_D$, for the SARS-CoV-2 N-protein and the S-protein on the Antibody/N$_3$-CNM and Antibody/N$_3$-CNM/graphene functionalized Au sensor slides measured with SPR.

|  | $k_a\ (Ms)^{-1}$ | $k_d\ (s^{-1})$ | $K_D\ (nM)$ |
|---|---|---|---|
| **Antibody/N$_3$-CNM** | | | |
| **N-protein** | $3.0 \pm 0.1 \times 10^6$ | $1.7 \pm 0.1 \times 10^{-3}$ | $0.57 \pm 0.05$ |
| **S-protein** | $1.24 \pm 0.13 \times 10^6$ | $2.79 \pm 0.10 \times 10^{-3}$ | $2.25 \pm 0.31$ |
| **Antibody/N$_3$-CNM/graphene** | | | |
| **N-protein** | $1.0 \pm 0.1 \times 10^4$ | $8.1 \pm 4.0 \times 10^{-6}$ | $0.8 \pm 0.5$ |
| **S-protein** | $5.1 \pm 0.1 \times 10^5$ | $6.2 \pm 0.5 \times 10^{-4}$ | $1.2 \pm 0.5$ |

## 4. Charge transfer model for the GFET response

The charge-transfer model used to interpret the GFET response is adapted from Ref.[6] In contrast to conventional analyses that treat the sensor response in a single buffer condition, we use the model here to explicitly analyze the responses for two different buffer conditions: (1) during target binding to the surface (which occurs in physiological ionic strength), and (2) electrical read-out of the bound charge (performed in an extremely low ionic strength). This two-step approach prevents the conflation of two distinct effects on the charge neutrality point of the GFET: from the antibody's intrinsic charge with the sensor offset potential, allowing each contribution to be evaluated separately. The GFET/electrolyte interface is described as a system of three capacitively coupled layers – (i) the graphene channel, (ii) the CNM + antibody layer, and (iii) the electrical double layer (EDL) in solution – which together must satisfy overall charge neutrality (see Figure S8 for a schematic illustration of these layers, where the charges on graphene, the surface, and in the EDL are denoted $n_{gr}$, $n_s$, and $n_{GC}$, respectively). The effective thickness of the surface adlayer (CNM + antibody) is ~5.4 nm, corresponding to the size of an IgG antibody atop the ~1 nm CNM film.[4] Below we outline the modeling procedure and key assumptions and then discuss the fitted parameters in context.

We apply the model in several steps to compute the expected $V_{CNP}$ shift ($\Delta V_{CNP} = V_{CNP\,[Signal]} - V_{CNP\,[Reference]}$) upon target binding:

**Step 1: Calibration in 1X Buffer (Offset Potential):** We first calibrate the model using measurements in undiluted physiological buffer (1X PBS-P, ionic strength ~166 mM) before any target is introduced. Under these conditions, the screening length is only a few nanometers, so the charge of the immobilized antibodies is effectively screened and does not contribute to the $V_{CNP}$ shift. The GFET's observed $V_{CNP}$ in 1X buffer is therefore governed by the intrinsic surface charge of the graphene/CNM and any fixed charges on the surface, $n_0 = 0.10 \pm 0.02\ C/m^2$ (e.g. from functional groups)[7]. Mathematically this corresponds to $n_S = n_0 + n_{AB} + n_f = n_0$ in the model, where $n_0$ is the surface charge density in the absence of antibodies and target proteins bound to the surface, $n_{AB}$ is the surface charge density of the antibodies, and $n_f$ is the (functional) surface charge of the adsorbed target proteins. We adjust the model's offset potential parameter so that the calculated $V_{CNP}$ matches the experimental value in 1X buffer. The offset potential is the only fitting parameter in this initial step and considers any potential differences between the signal and the reference device in the absence of any surface functionalization.

**Step 2: Antibody Charge in 0.0003X Buffer:** We calculate the antibody charge from the difference of the experimental $V_{CNP}$ values measured in the physiological buffer and the diluted buffer, $V_{CNP}(0\ aM, 1X\ PBS - P) - V_{CNP}(0\ aM, 0.0003X\ PBS - P)$, as we can assume that the antibody charge is completely screened in the physiological buffer solution, and completely unscreened in the diluted buffer solution. In the low-ionic-strength buffer, the Debye length ($\lambda_D \approx$ 40 nm)[6] is larger than the size of the antibodies (~15 nm).[4] Experimentally, switching from 1X to 0.0003X buffer causes a substantial $V_{CNP}$ shift (on the order of hundreds of millivolts). In our case, we observed a shift of ~290 mV. According to the model, ~150 mV of this shift is originating from ionic strength change alone. We attribute this extra ~140 mV shift to the de-shielding of the antibody's negative charges on the surface. Using the known offset potential and offset charge density $n_0$ we calculate the surface charge density $n_{AB}$ of the antibody layer that would produce the additional shift. The fitting yields $n_{AB} \approx -0.27$ C/m², which corresponds to roughly $-20$ elementary charges per antibody (consistent with the fact that IgG's antibody net charge at pH 7.4 can be negative). We fix this antibody charge density in the model for subsequent calculations. To accurately determine the hysteresis-corrected $V_{CNP}$, we calculate the average value measured in positive and negative sweeping directions. At an ionic strength of 55 µM we obtain $V_{CNP}(c_i = 55\ \mu M) = 300 \pm 10\ mV$ for the Ab(S) and $V_{CNP}(c_i = 55\ \mu M) = 290 \pm 10\ mV$ for the a-N-protein antibody, where $c_i$ is the ionic strength of the solution.

**Step 3: Langmuir–Freundlich Isotherm (Target Binding in 1X buffer and Target Response Measurement in 0.0003X Buffer):** With the sensor baseline established, we model the binding of target proteins to the antibodies under physiological ionic strength using a Langmuir–Freundlich adsorption isotherm.[6]

$$n_f = \frac{N_f z_e}{1+10^{m_f(\log\log K_D\ -\log\log c)}\,e^{m_f \beta z_{eff} \Psi_S}},$$

where $N_f$ is the number density of available specific binding sites for the target, $z_{eff} = ze^{-d/\lambda_D}$ is the effective charge of the protein during the binding event, $\lambda_D$ is the Debye length, $d$ is the effective distance between the charges during a binding event, $m_f$ is the heterogeneity parameter, $K_D$ is the constant for the binding of target molecule to the antibody, $c$ is the concentration of the target in solution, $\beta = 1/k_B T$; $k_B$ is the Boltzmann constant; $T$ is temperature and $\Psi_S$ is the surface potential. This isotherm extends the classic Langmuir model by introducing a heterogeneity parameter ($m_f$) to allow for a distribution of binding affinities (with $m_f = 1$ reducing to the homogeneous Langmuir case). To account for electrostatic interactions between charged targets on the surface, a Boltzmann factor is included in the isotherm.[6] Specifically, if a target protein of effective charge $z_{eff}$ binds at a surface potential $\psi_s$, the local concentration (and thus binding probability) is modified by a factor $exp(-z_{eff} e \Psi_s / k_B T)$, where $e$ is the elementary charge and $k_B T$ is the thermal energy. In practice, this means that as targets bind and impart charge to the surface, further binding can be hindered (for like-charged targets being repelled by a similarly charged surface) or enhanced (for oppositely charged targets). We use the equilibrium dissociation constant ($K_D$) measured by SPR for the target–antibody interaction to fix the intrinsic binding affinity in the model (for N-protein, $K_D \approx 0.57$ nM; for S-protein, $K_D \approx 0.022$ nM)[4]. The unknowns at this stage are the heterogeneity factor $m_f$, the total density of binding sites $n_{f,max}$ (which should be on the order of $N_{AB}$, scaled by the number of epitopes per antibody), and the effective interaction distance $d$ (which enters into the Boltzmann factor via the Debye length, determining how strongly surface potential influences binding). We then calculate the corresponding shift in $V_{CNP}$ when the device is measured in diluted buffer. Here we use the charge-neutrality condition in the three-layer system. As noted above, in 0.0003X buffer we treat the Debye screening as negligible ($\lambda_D$ ~40 nm ≫ device dimensions), so the full charge of each target contributes. The presence of the pre-existing antibody layer charge (determined in step 2) and the GFET's intrinsic charge (from step 1) are both included in this calculation. In physiological buffer, this charge is strongly screened – effectively, the protein's charge influence is confined to a Debye length (~0.7 nm in 1X PBS-P). We account for this by using a reduced effective charge $z_{eff}$ in the Langmuir–Freundlich model (step 3) that reflects the screened interaction during the binding process. For this screened charge $n_{f,screened}$, we calculate the ratio of occupied adsorption sites to the total number of adsorption sites, $j(c) = \frac{n_{f,screnned}}{N_f z_{eff}}$. In contrast, in diluted buffer, screening is negligible, so each bound protein contributes its full unscreened charge $z$ (taken to be the net charge at pH 7.4) to the surface potential. The outcome, which is $\Delta V_{CNP}$ as a function of target concentration, is then calculated by keeping $j(c)$ constant but using $n_f = j(c)N_f z$, where $z$ is the unscreened charge of the target, resulting in the required value of $\Delta V_{CNP}$. The change of $V_{CNP}$, the surface charge density and the potential due to target binding is shown in Figure S9 and Table S2.

**Table S2.** Surface Charge Density and Potential at Various Planes for Different S-Protein Concentrations. This table displays the calculated surface charge density n and potential $\Psi$ at three critical planes within the device: the outer Helmholtz plane (OHP), the functional (surface) plane (S), and the graphene plane (Gr) for varying target concentrations, ranging from 1 aM to 1 µM. This table offers a detailed view of the model's output, crucial for understanding the interactions within the graphene FET sensor at different target concentrations.

| $V_{EG} = 0$ | $c_{BIO} = 10^{-18}$ M | $c_{BIO} = 10^{-12}$ M | $c_{BIO} = 10^{-6}$ M |
|---|---|---|---|
| $\Psi_{OHP}$ [V] | -0.328 | -0.307 | -0.307 |
| $\Psi_s$ [V] | -0.437 | -0.380 | -0.378 |
| $\Psi_{gr}$ [V] | 0.140 | 0.140 | 0.140 |
| $n_{GC}$ [C/m²] | 0.262 | 0.174 | 0.173 |
| $n_s$ [C/m²] | -0.271 | -0.181 | -0.179 |

**Step 4: Fitting and Parameter Extraction:** We adjust the model's free parameters – primarily the effective distance $d$, binding site density $n_{f,max} = N_f z$, and heterogeneity parameter $m_f$ introduced in step 3 – to achieve the best fit between the predicted $\Delta V_{CNP}$ vs. concentration curve and the experimental data (for each target protein). The antibody charge density and offset potential determined in steps 1–2 are held fixed during this fitting. We perform this optimization separately for the N-protein and S-protein datasets (using their respective $K_D$ values and protein charges). The fitting process yields a specific set of parameters ($d$, $n_{f,max}$, $m_f$) for each target that best explain our observations, see Table S3.

**Table S3.** Summary of modeled parameters surface charge density of the GFET.

|  | $n_0$ [$C/m^2$] | $n_{AB}$ [$C/m^2$] | $n_{f,max}$ [$C/m^2$] | $d$ [$nm$] | $m_f$ |
|---|---|---|---|---|---|
| *N-protein* | $0.10 \pm 0.02$ | $-0.26 \pm 0.13$ | $0.075 \pm 0.036$ | 1.7 | 1.0 |
| *S-protein* | $0.10 \pm 0.02$ | $-0.28 \pm 0.14$ | $0.11 \pm 0.05$ | 1.85 | 0.5 |

The heterogeneity parameter $m_f = 1$ for the N-protein indicates homogeneous Langmuir-type binding, whereas $m_f = 0.5$ for the S-protein suggests a broader distribution of binding affinities, potentially arising from varying epitope orientations or antibody binding configurations on the surface.

The effective interaction distance $d$ reflects the mean spacing between antibodies on the surface. Based on the average antibody surface coverages of approximately $0.10 \pm 0.05$ nm$^{-2}$ and $0.08 \pm 0.04$ nm$^{-2}$ for the S- and N-protein antibodies, respectively, individual antibodies are spaced approximately 3 nm apart, which is consistent with the fitted interaction distances of $d \approx 1.7\text{-}1.85$ nm.

The limit of detection (LOD) was determined from the model as the lowest analyte concentration at which the predicted sensor response ($\Delta V_{CNP}$) becomes distinguishable from the baseline noise. The baseline noise was quantified as one standard deviation (1σ) of the experimental differential measurement for the blank sample (0 M target), derived from the data presented in Figure 4. Applying this 1σ criterion, the model estimates the LOD to be approximately 5 aM for the S-protein and 95 aM for the N-protein, consistent with the experimental observations.

## 5. Hysteresis suppression through differential readout in GFET arrays

At low ionic strength, GFETs are susceptible to signal instabilities arising from hysteresis in their transfer curves, which originates from the interplay between the electrolyte-gate capacitance and parasitic capacitances induced by unintended impurity doping. This behavior complicates reliable sensing in diluted buffer conditions. To mitigate this effect, the differential responses of all signal–reference combinations (Figure S7a) were evaluated using the last two data points acquired in the diluted buffer as the reporting values (Figure S10). In addition, all signal and reference devices were measured simultaneously, ensuring identical gate-voltage sweep histories across the array. As a result, hysteresis-related artifacts are largely suppressed, enabling reliable extraction of differential sensorgrams (see Figure S10).

## 6. Calculation of the charge of the N-protein

We use an N-protein from the Severe acute respiratory syndrome coronavirus 2 isolate Wuhan-Hu-1 Ref.[8]. The sequence of the N-protein is

SDNGPQNQRNAPRITFGGPSDSTGSNQNGERSGARSKQRRPQGLPNNTASWFTALTQHG

KEDLKFPRGQGVPINTNSSPDDQIGYYRRATRRIRGGDGKMKDLSPRWYFYYLGTGPEA

GLPYGANKDGIIWVATEGALNTPKDHIGTRNPANNAAIVLQLPQGTTLPKGFYAEGSRGGSQASSRSSSRSRNSSRNSTPGSSRGTSPARMAGNGGDAALALLLLDRLNQLESKMSGKGQQQQGQTVTKKSAAEASKKPRQKRTATKAYNVTQAFGRRGPEQTQGNFGDQELIRQGTDYKHWPQIAQFAPSASAFFGMSRIGMEVTPSGTWLTYTGAIKLDDKDPNFKDQVILLNKHIDAYKTFPPTEPKKDKKKKADETQALPQRQKKQQTVTLLPAADLDDFSKQLQQSMSSADSTQASAWSHPQFEK.

We used the platform Prot pi to calculate the charge of the protein, the charge at a pH value of 7.4 is +23.3e (Ref. [9, 10]). It is important to note that this value is an approximation, as the protein is biologically produced and undergoes processing (e.g., purification and potential cleavage) by the manufacturer. The exact final sequence is difficult to ascertain, and even minor deviations from the reference could lead to a substantial difference in the actual net charge. Nevertheless, this theoretical value is in qualitative agreement with experimental observations of the N-protein antigen layer, which also confirms a highly positive charge.

## 7. Data acquisition for the GFET sensor response

The data acquisition software was developed in LabVIEW 2020. The program controlled the drain-source voltages of the 15 GFETs as well as the gate potential, while simultaneously recording the corresponding drain–source currents to obtain the transfer curves of all devices. The charge neutrality point ($V_{CNP}$) was determined in real time by fitting the transfer curves with a third-order polynomial function implemented in LabVIEW, and the extracted values were used to generate the $V_{CNP}$ sensorgram.

## 8. Comparison of Readout Architecture for GFET Arrays

Here we compare different readout techniques for GFETs.

### 8.1. Single-Channel Sequential Measurement

This method for GFET characterization involves the use of a single, high-precision, laboratory-grade instrument, typically a benchtop Source-Measure Unit (SMU), to measure one device at a time. The SMU is manually or semi-automatically connected to the source, drain, and gate terminals of an individual GFET, and a full electrical characterization (e.g., transfer curves, output curves) is performed.[6, 11] The measurement process is repeated for each device in the array. The high accuracy, low noise, and broad dynamic range of a research-grade SMU ensure that the measured data reflects the true behavior of the GFET with minimal instrumentation-induced artifacts. However, its application to arrays is severely limited by a low throughput. Characterizing an array of hundreds of devices in this manner is very time-consuming, rendering it impractical for any application requiring real-time data acquisition with large device arrays.

### 8.2. Multiplexed (Time-Division Multiplexed) Systems

The most frequently used architecture for reading out GFET arrays without manual switching between devices involves the usage of a single source meter in combination with a fast electronic switch that enables to select between multiple devices. This approach addresses the throughput limitation of single-channel measurements by automating the switching process between sensors. The core principle of this architecture is Time-Division Multiplexing (TDM), a technique in which a limited number of measurement channels are sequentially connected to a much larger number of sensors through a network of high-speed electronic switches, known as multiplexers (MUX) and demultiplexers (DEMUX).[12, 13] A central control unit, such as a microcontroller, directs the switches to connect the measurement instrument(s) to a specific sensor, acquires the data, and then rapidly switches to the next sensor in the sequence. This process is inherently sequential; measurements are taken one after another in rapid succession, not at the same time.

**PCB-Based Multiplexing:** This is the most common custom-built approach, where the multiplexing switches, signal amplification circuitry, and control logic are integrated onto a custom-designed printed circuit board (PCB).[14] The PCB interfaces with the GFET array chip and a computer or portable readout unit. This approach has been successfully used to automate measurements for arrays ranging from a few devices to over 200,[14] see Table S4.

**Monolithic CMOS Multiplexing:** This represents the most advanced and scalable implementation of TDM. In this architecture, the multiplexing circuitry, including switches and sometimes even the analog front-end electronics, is fabricated directly onto the same silicon chip that serves as the substrate for the GFETs.[15] This monolithic integration enables the creation of extremely dense sensor arrays, with demonstrations of over 500 GFETs on a single chip, and offers a clear path toward arrays with thousands or even millions of sensors. This approach enables the development of low-cost high-density biosensing chips.

### 8.3. Parallel Multi-Channel Systems

A distinct, though less common, architecture is the true parallel multi-channel system. This approach fundamentally differs from TDM by employing one physically separate and independently operating measurement channel for each sensor in the array. In the context of GFETs, this would be realized with an array of independent Source-Measure Units (SMUs), with each SMU dedicated to a single GFET.

We use this architecture here, as it enables truly simultaneous data acquisition with very high temporal resolution and accuracy. Because each sensor has its own dedicated sourcing and measurement hardware, all sensors in the array can be stimulated and measured at the exact same instant in time. In data acquisition engineering, this architecture is also referred to as phase-synchronous measurement, as it ensures that there is a stable and known phase relationship (ideally zero) between the data points acquired from different channels. The complete elimination of time delay between channels maximizes the accuracy of data for dynamic measurements where the relative timing of events is critical.

A common implementation of a TDM system uses a single, shared instrumentation amplifier to read the current from all sensors in the array. This creates an unavoidable trade-off in all TDM systems between measurement speed and accuracy: to increase accuracy, one must increase the dwell time on each channel to allow for full settling, which in turn reduces the overall throughput. A parallel system with a dedicated amplifier for each channel completely circumvents this issue. The increase of the number of devices does not increase the measurement time as each device has its dedicated source meter. Thus, when multiple source meters are used to acquire data in parallel, their simultaneous measurements reduce the overall noise and improve the signal-to-noise ratio for a measurement performed within a defined time window. Thus, high temporal resolution can be achieved even for larger device arrays. In addition, a true parallel system, built with physically separate and properly shielded SMUs for each channel, inherently eliminates shared electrical paths, thereby providing maximum isolation and minimizing the risk of inter-channel crosstalk. A comparative analysis of the usage of these readout techniques in the literature is presented in Table S4.

**Table S4**. Summary of literature analysis of GFET readout architectures. Abbreviations used: SMU: source measure unit, MUX: multiplexer, PCB: printed circuit board, Measurement method "transfer characteristic": acquiring the complete transfer characteristic for each measurement, "I/V spot": initial measurement of the complete transfer characteristic, followed by tracking the sensor's response by monitoring the current at a single, fixed gate voltage. True differential" indicates that signal and reference devices are measured simultaneously at the exact same time point, enabling accurate correction for baseline drift and environmental effects.

| GFET | | SMU-channels | | MUX | measurement | | | parameter | | Ref |
|---|---|---|---|---|---|---|---|---|---|---|
| array size | count per chip | lab device | app.-spec. PCB | type | sequence | method | CNP | true differential | | |
| single device | 1 | 1 | - | - | - | transfer curve | yes | no | | [11] |
| small array (2-10) | 5 | 1 | - | automated | serial | transfer curve | yes | no | | [12], [13] |
| medium array (11-50) | 12 | - | 1 | automated | serial | transfer curve | yes | no | | [16] |
| | 12 | - | 1 | automated | serial | I/V spot | no | no | | [17] |
| | 15 | 1 | - | manual | serial | transfer curve | yes | no | | [6] |
| | 20 | - | 1 | automated | serial | I/V spot | no | no | | [18] |
| large array (>50) | 256 | - | 1 | automated | serial | I/V spot | no | no | | [14] |
| | 512 | 1 | - | automated | serial | tranfer curve | yes | no | | [15] |
| medium array (11-50) | **15** | **-** | **15** | **-** | **parallel** | **transfer curve** | **yes** | **yes** | | **this work** |

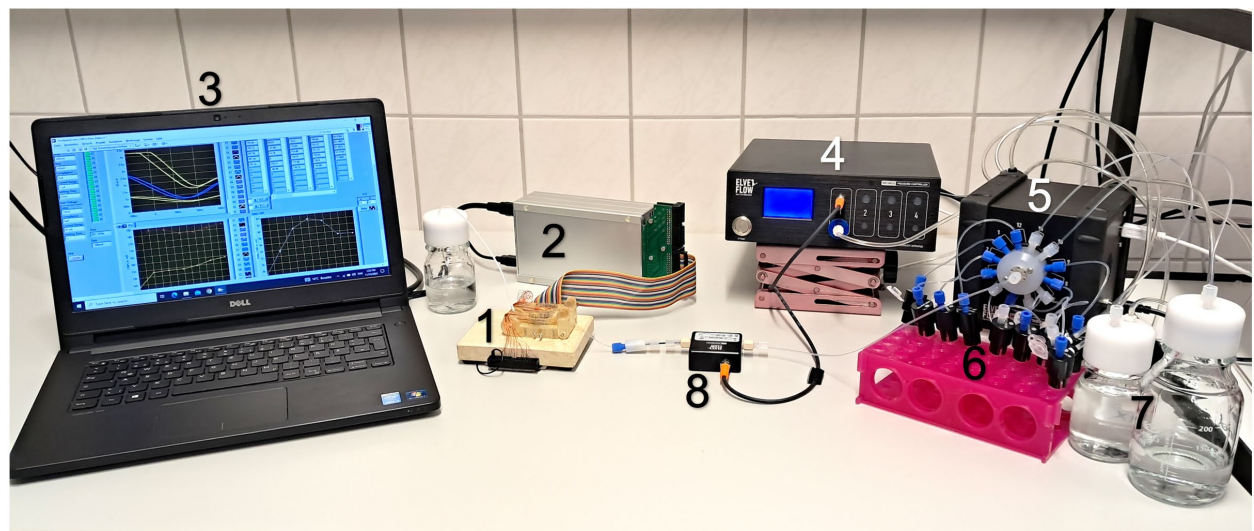

**Figure S1**. Experimental setup: 1) Microfluidic cell with a GFET chip. 2) 16-channel source measure unit. 3) Software on a laptop showing the graphical user interface. 4) Pressure regulator. 5) Fluidic distributor with 12 channels. 6) Antigen solution reservoirs. 7) Physiological and diluted buffers (1x and 0.0003X PBS-P). 8) Flow sensor.

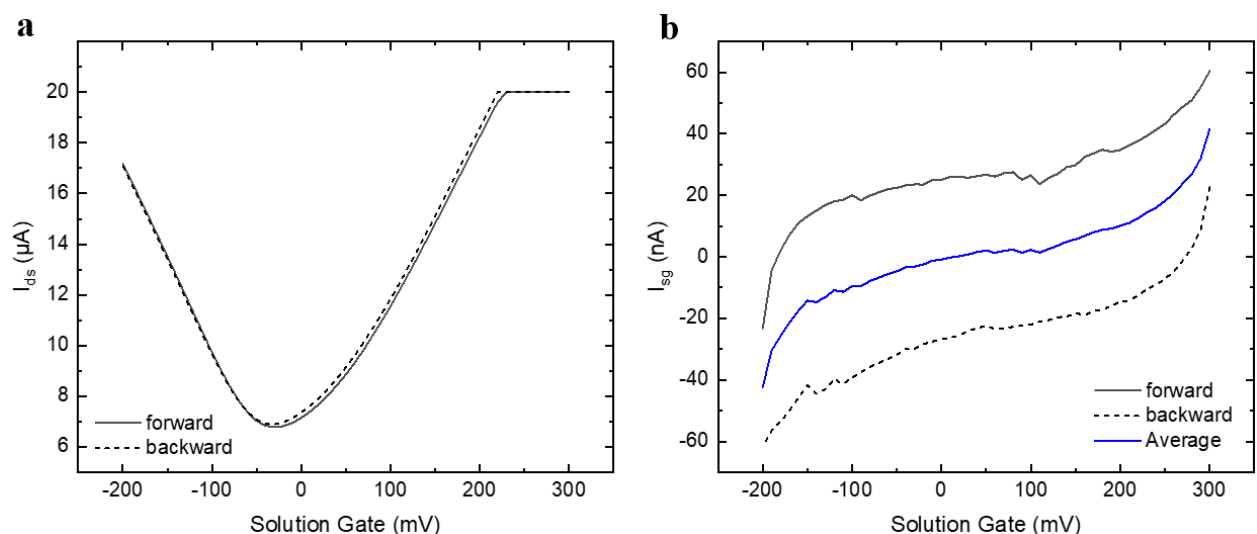

**Figure S2**. Gate-source leakage current measurement. a) Transfer characteristic curve of a typical device measured in 1x PBS-P ($V_{ds}$= 10 mV). The forward and backward gate sweepings are indicated with solid and dashed curves respectively. b) The gate-source current ($I_{sg}$) measured over the corresponding gate cycles. The blue-colored solid curve represents the hysteresis-corrected leakage current calculated from the average of the forward and backward gate sweepings leakage current measurements.

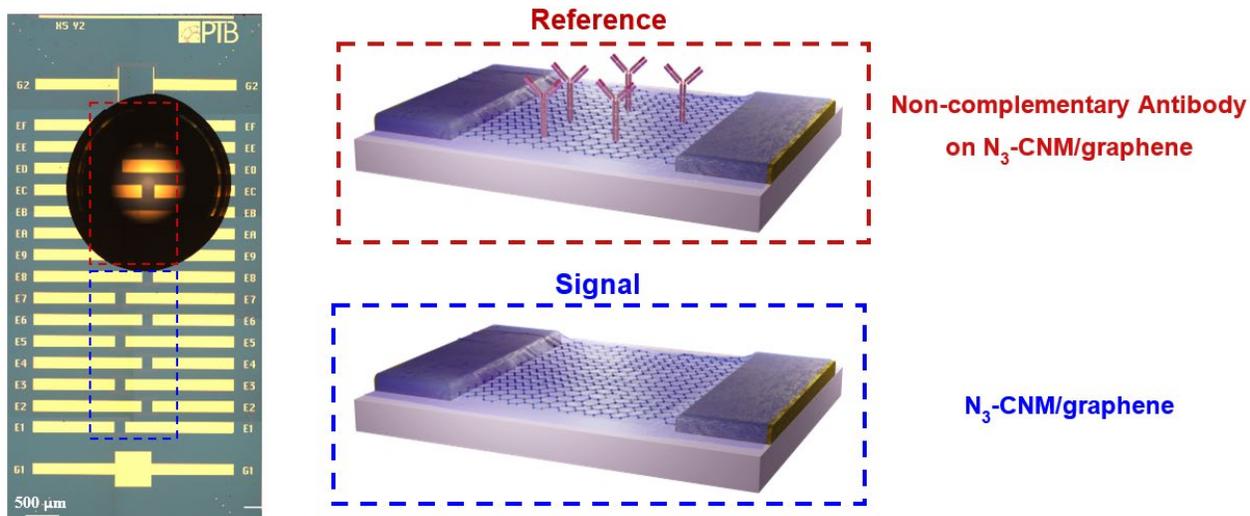

**Figure S3**. Functionalization of a GFET chip by drop casting (*ex situ*) before *in situ* Ab$_{(S)}$ functionalization on N$_3$-CNM/graphene surface. The Ab$_{(Ref)}$-functionalized GFETs (E10-E15) serve as reference devices during the real-time monitoring of the Ab$_{(S)}$ immobilization on signal devices and the real-time monitoring of the target response. The functionalization of reference devices involves drop-casting the Ab$_{(Ref)}$ (100 µg/ml antibody solution in 10 mM sodium acetate at pH 4.5) onto selected N$_3$-CNM/GFET devices (E10-E15) while keeping the rest of the devices (E01-E05) unaffected. After applying the solution, these devices are maintained in a humid environment for one hour, followed by rinsing the chip with ultra-pure water and drying it in ambient condition.

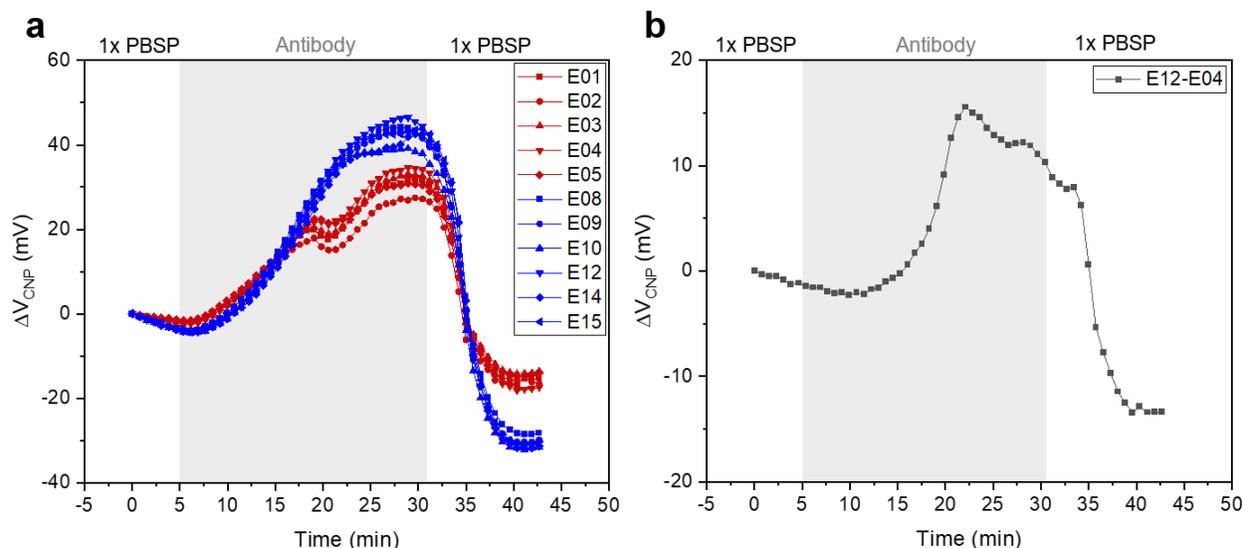

**Figure S4.** Comparison of SARS-CoV-2 antibody immobilization on $N_3$-CNM/graphene versus bare graphene using a half CNM-covered GFET chip a) Time dependent $\Delta V_{CNP}(t)$ values ($\Delta V_{CNP}(t) = V_{CNP}(t) - V_{CNP}(t=0)$) sensorgrams of bare graphene devices (E01-E05, red-colored symbols) and $N_3$-CNM/graphene devices (E08-E15, blue-colored symbols). The measurement buffer is physiological buffer (1X PBS-P). The injection time of the $Ab_{(S)}$ is highlighted with grey (100 µg/mL in 10 mM sodium acetate, pH 4.5, 5 µL/min). During the antibody injection (grey highlighted), the distinct $V_{CNP}$ responses of signal and reference devices can be attributed to the antibodies' conformational changes while adsorbing on bare graphene. These trends were reproducible across three independently fabricated GFET chips. b) The corresponding differential sensorgram of a typical signal-reference pair shown in a.

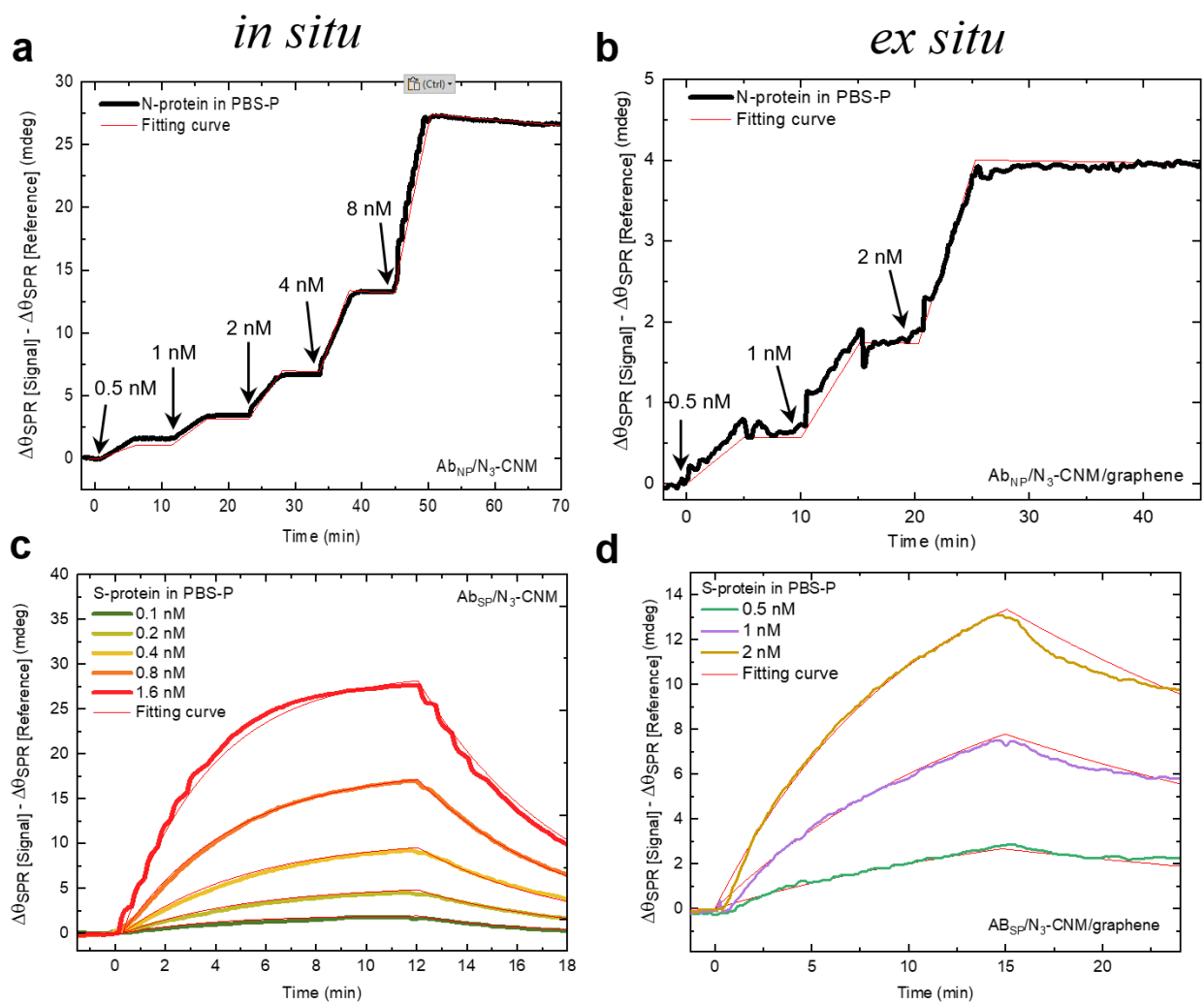

**Figure S5**. Analysis of the binding of SARS-CoV-2 N-protein (a, b) and S-protein (c, d) to the Antibody/$N_3$-CNM (a,c) and Antibody/$N_3$-CNM/graphene functionalized Au sensor slides with SPR. Real-time referenced resonance angle shift during sequential N- and S-protein injections into signal and reference channels. The signal channels were functionalized with $Ab_{(N)}$ or $Ab_{(S)}$ for the detection of the N- and S-protein, respectively. The reference channel was functionalized with $Ab_{(Ref)}$. Antibody functionalization was done *in situ* (a, c) or *ex situ via* drop-casting (b, d). Black arrows in panels a and b mark the injection times for each N-protein sample during the single cycle kinetic measurements. For the S-protein, the surface was regenerated between the injection of the samples.

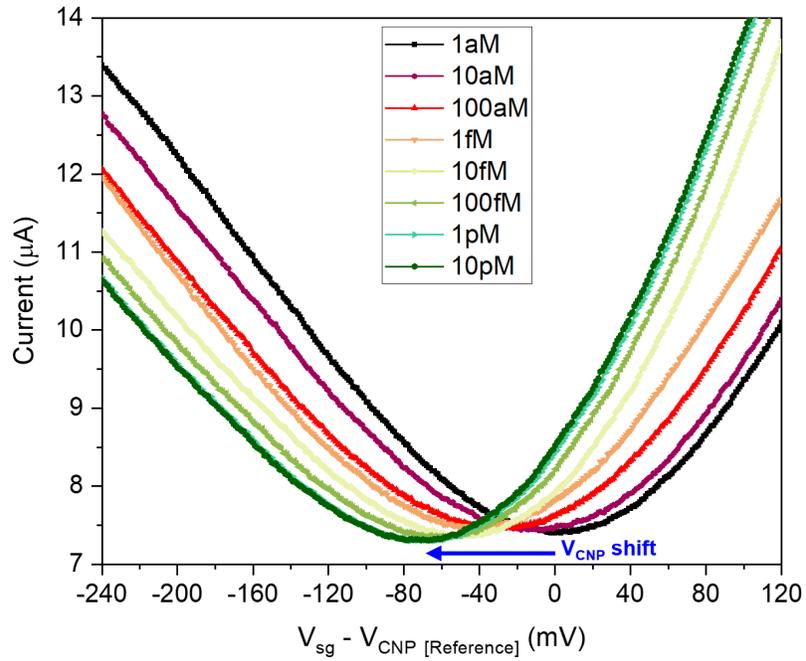

**Figure S6**. Reference-corrected transfer curves of a typical signal device (Ab$_{(S)}$-functionalized) measured in the diluted buffer (0.0003X PBS-P, $V_{ds}$= 10 mV) after sequential S-protein concentration injection (1 aM-10 pM) in physiological buffer (1X PBS-P).

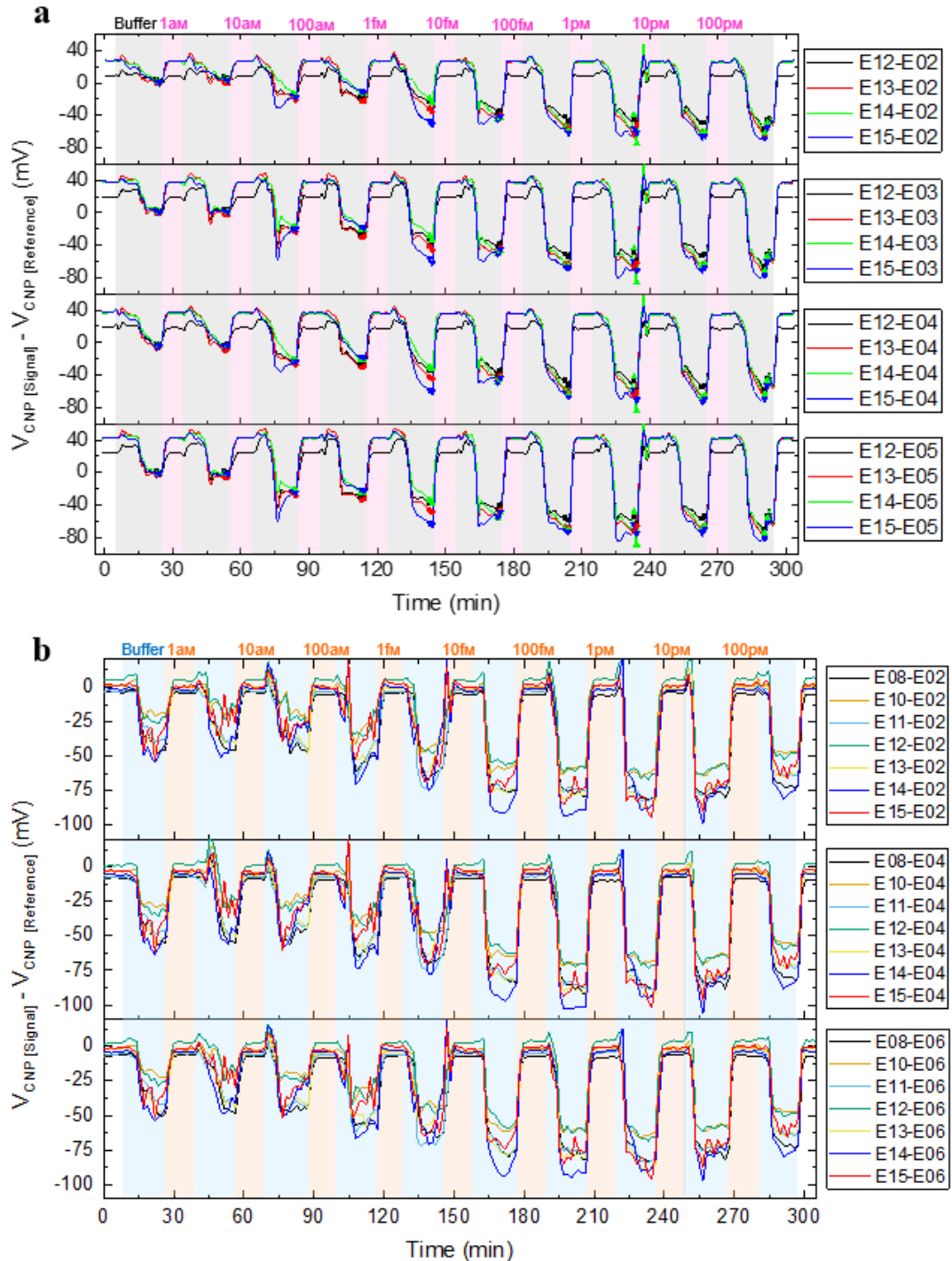

**Figure S7**. Differential sensorgrams of two pairs of signal-reference as indicated in the legend: a) Concentration series of S-protein (16 pairs) and b) For N-protein (21 pairs). Reporting points are denoted by colored symbols corresponding to the graph's color. Grey shadings indicate the injection phases of the measurement buffer (0.0003X PBS-P), while pink and orange shades in a and b, respectively, highlight the injection periods for S- and N-protein samples.

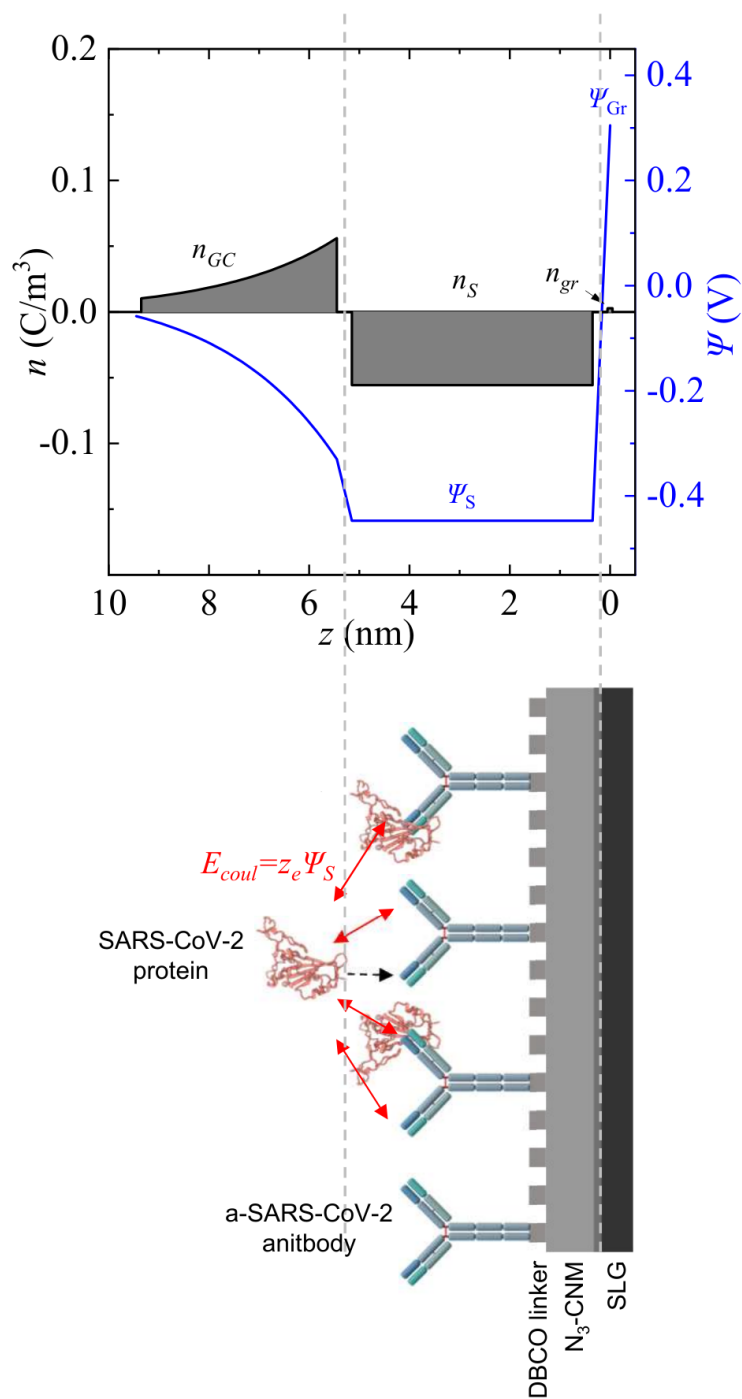

**Figure S8**. Schematic representation of the electrical potential $\Psi$ and charge density n in the Gouy-Chapman layer (GC), on the surface (S) and in graphene (Gr). The positions of the target protein, the antibody, the DBCO-linker, the $N_3$-CNM and SLG are indicated. $E_{coul}$ is the Coulomb interaction and $z_e$ is the effective charge of the target upon binding.

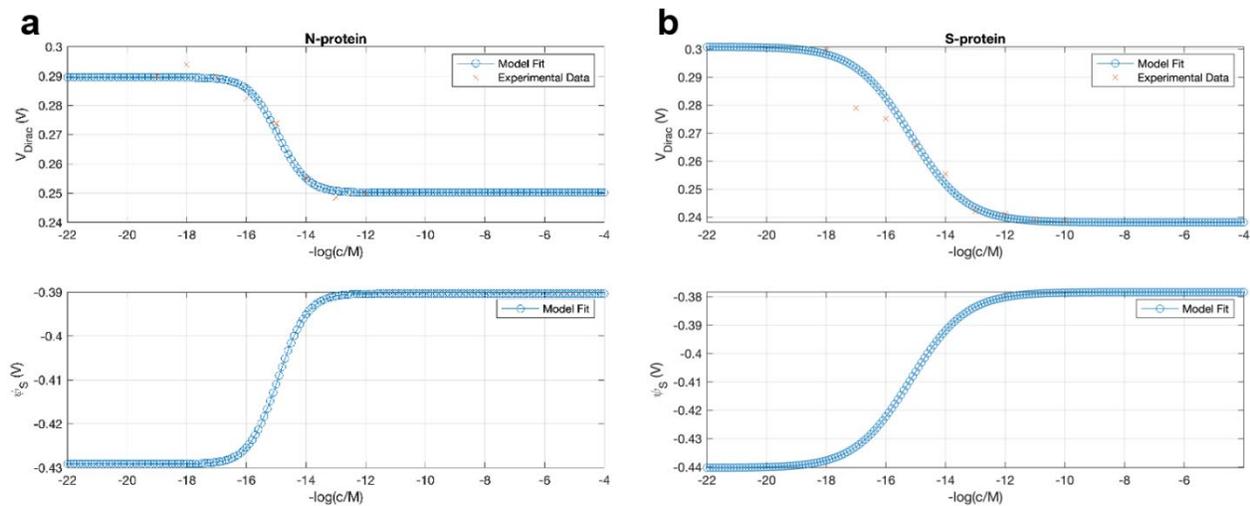

**Figure S9**. Modeling of the GFET biosensor response for (a) SARS-CoV-2 N-protein and (b) SARS-CoV-2 S-protein. Top: Model results (lines) and experimental data (points) for the sensor response, $V_{CNP}$, as a function of the target concentration. Bottom: Model results for the surface potential $\Psi_S$ for various concentrations of the target in solution.

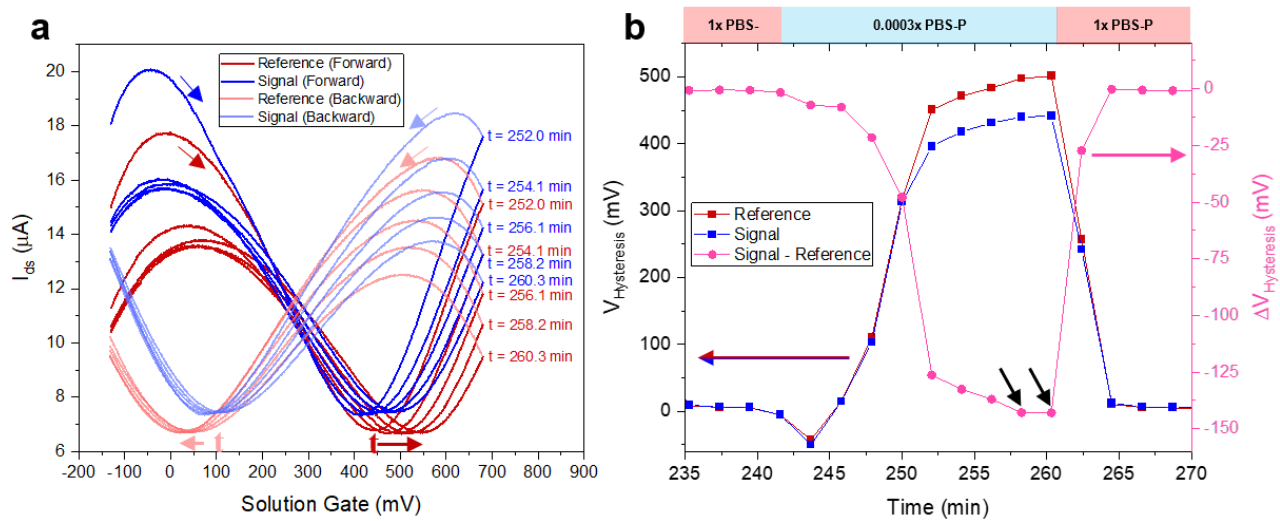

**Figure S10**. a) Transfer characteristics of a signal device (blue curves) and a reference device (red curves), captured at specific time corresponding to the recorded data points in the diluted buffer (0.0003X PBS-P). The solid and semi-transparent curves represent forward and backward sweeps, respectively, as indicated by the direction of the arrows. b) The $V_{CNP}$ hysteresis of the corresponding signal (blue-colored symbols) and reference (red-colored symbols) devices (left Y-axis), along with their relative hysteresis ($ΔV_{CNP}$, pink-colored symbol, right Y-axis) in a dual Y-axis graph. The data points are collected during a cycle transitioning from the physiological buffer to the diluted buffer, followed by reverting to the physiological buffer. The black arrows mark the data points at which the relative hysteresis of signal and reference devices become minimal, thereby selecting these data points as the reporting points in Figure S7.

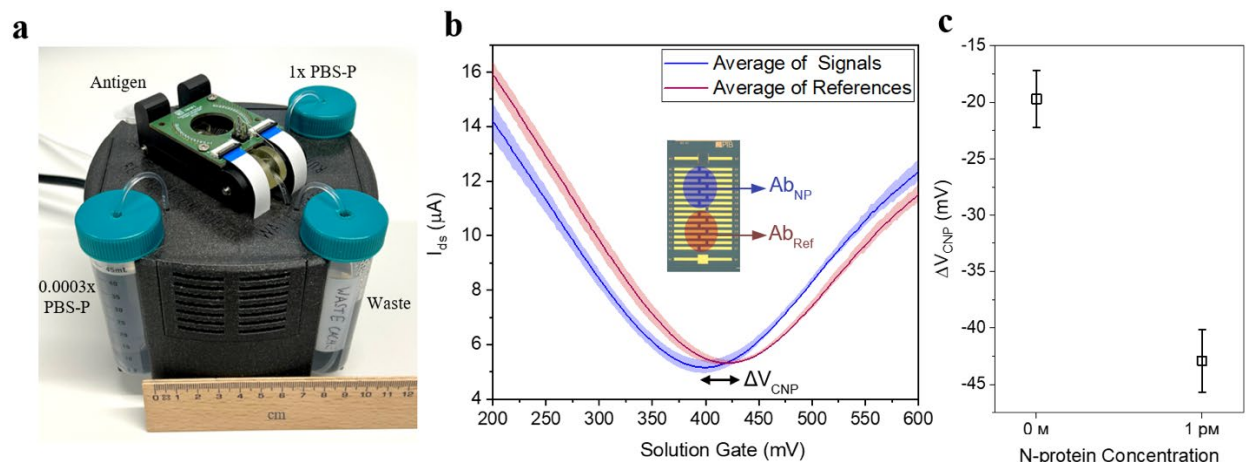

**Figure S11**. a) Photograph of the portable sensing platform. b) Mean (solid) and standard deviation (filled area) of transfer curves of three signal devices (blue curve) and three reference devices (red curve) in 0.0003X PBS-P buffer. c) Average of $\Delta V_{CNP}$ (signals – references) after injection of 0M and 1pM N-protein.

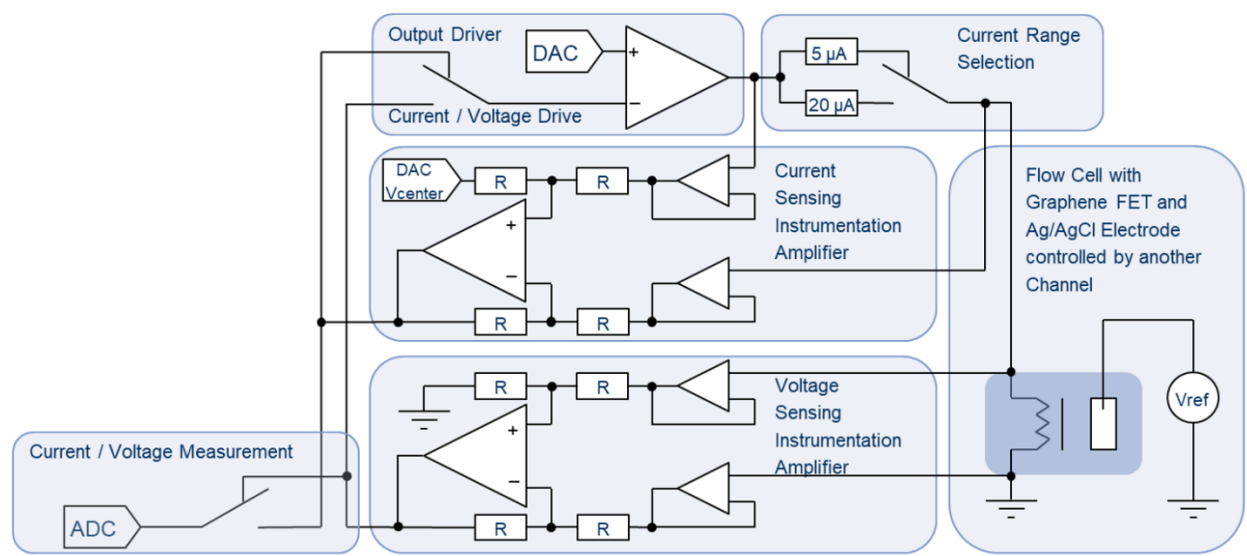

**Figure S12**. Circuit topology of a channel in the source meter electronic measurement environment.